%% file: main.tex
\documentclass[fleqn,10pt,twocolumn]{wlscirep}

\usepackage[T1]{fontenc}
\usepackage{amsthm,bm}
\usepackage{tikz,graphicx,dcolumn}
\usepackage{amssymb}
\usepackage{color}
\usetikzlibrary{shapes}
\usepackage{standalone}
\usepackage{makecell}
\usepackage[utf8]{inputenc}
\usepackage[justification=justified]{caption}
\usepackage{comment}
\usepackage{hyperref}

\DeclareRobustCommand\markerone{{\tikz{\node[draw,scale=0.5,circle,fill=none,black!20!green](){};}}}
\DeclareRobustCommand\markertwo{{\tikz{\node[draw,scale=0.5,diamond,fill=none,black!20!blue](){};}}}
\DeclareRobustCommand{\markerthree}{{\tikz{\node[draw,thick,scale=0.9,fill=none,black!20!red](){};}}}

\DeclareRobustCommand{\blackline}{\raisebox{2pt}{\tikz{\draw[thick,dotted](0,0)--(5mm,0);}}}

\title{Quantum-accurate magneto-elastic predictions with classical spin-lattice dynamics}

\author[1]{Svetoslav Nikolov}
\author[1]{Mitchell A. Wood}
\author[2]{Attila Cangi}
\author[3,4]{Jean-Bernard Maillet}
\author[5]{Mihai-Cosmin Marinica}
\author[1]{Aidan P. Thompson}
\author[6]{Michael P. Desjarlais}
\author[1,*]{Julien Tranchida}

\affil[1]{Computational Multiscale Department, Sandia National Laboratories,
P.O. Box 5800, MS 1322, Albuquerque, NM 87185}
\affil[2]{Center for Advanced Systems Understanding (CASUS),
Helmholtz-Zentrum Dresden-Rossendorf, 02826 Görlitz,
Germany}
\affil[3]{CEA - DAM, DIF, Arpajon Cedex F-91297, France}
\affil[4]{Université Paris-Saclay, CEA, LMCE, 91680 Bruyères-le-Châtel, France}
\affil[5]{Université Paris-Saclay, CEA, Service de Recherches de Métallurgie Physique, Gif-sur-Yvette 91191, France}
\affil[6]{Sandia National Laboratories,
P.O. Box 5800, MS 1322, Albuquerque, NM 87185}

\affil[*]{jtranch@sandia.gov}

\begin{abstract}

A data-driven framework is presented for building magneto-elastic machine-learning interatomic potentials (ML-IAPs) for large-scale spin-lattice dynamics simulations. 
The magneto-elastic ML-IAPs are constructed by coupling a collective atomic spin model with an ML-IAP. Together they represent a potential energy surface from which the mechanical forces on the atoms and the precession dynamics of the atomic spins are computed.
Both the atomic spin model and the ML-IAP are parametrized on data from first-principles calculations.  
We demonstrate the efficacy of our data-driven framework across magneto-structural phase transitions by generating a magneto-elastic ML-IAP for $\alpha$-iron. 
The combined potential energy surface yields excellent agreement with first-principles magneto-elastic calculations and quantitative predictions of diverse materials properties including bulk modulus, magnetization, and specific heat across the ferromagnetic-paramagnetic phase transition.
\end{abstract}

\usepackage{lineno}

\begin{document}

\flushbottom
\maketitle
%
%


\section*{Introduction}
Magnetism strongly influences thermomechanical properties in a large variety of materials, such as single-element magnetic metals \cite{tatsumoto1959temperature,bahl2009effect}, steels \cite{tavares2000magnetic}, high-entropy alloys~\cite{huang2018mapping,rao2020unveiling}, nuclear fuels such as uranium dioxide~\cite{jaime2017piezomagnetism},  magnetic oxides~\cite{nussle2019dynamic,lejman2019magnetoelastic}, and numerous other classes of functional materials \cite{patrick2020spin}. 
Despite the critical role of magnetism in the aforementioned materials classes, modeling efforts to study the interplay between structural and magnetic properties have been notably lacking.
Furthermore, there are unanswered scientific questions regarding the significance of magnetism in matter that is shock-compressed~\cite{GMVC1986:materials,surh2016magnetostructural} or exposed to strong electromagnetic fields such as in coherent lights sources\cite{Moses_NIF,Tschentscher_2017}, pulsed power and high magnetic fields facilities \cite{tan2014combined,gracia2020multicaloric}. 
Properties of interest include phase transitions, thermal stability of magnetic defects, magneto-mechanical couplings, but many of these subjects are challenging or prohibited by state of the art computational tools.

\begin{figure}[h!]
\centering
\includegraphics[width=0.95\columnwidth]{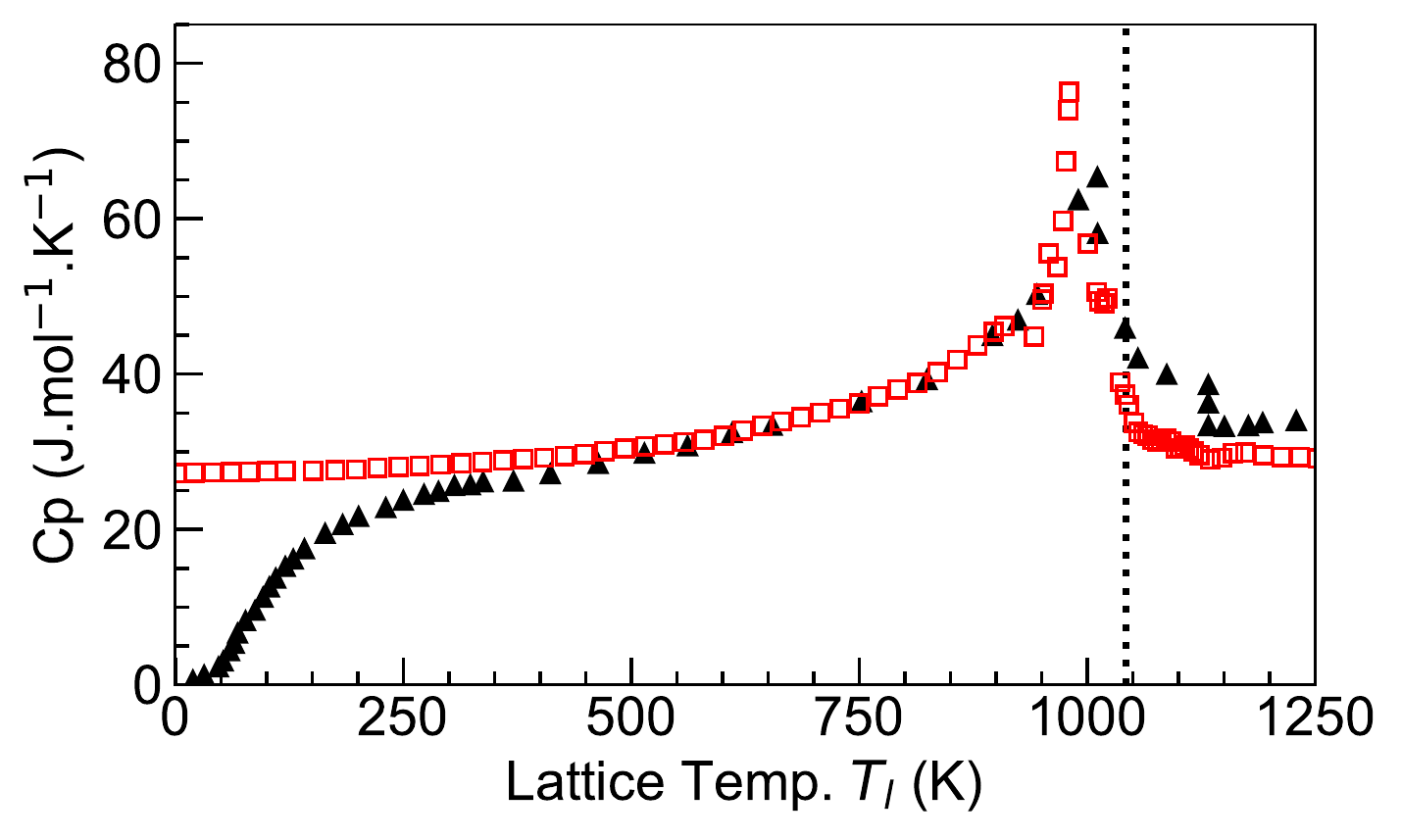}
\caption{Constant pressure heat capacity of $\alpha$-iron versus temperature. The black triangles denote experimental measurements~\cite{wallace1960specific,touloukian1970thermophysical}, the red squares our simulation results, and black dashed line indicates the experimental Curie transition temperature. This illustrates the well-known ferromagnetic-paramagnetic phase transition, where the heat capacity diverges at the Curie temperature.}
\label{fig:cp}
\end{figure}

A prime but simple example of the computational advance made herein is the heat-capacity of $\alpha$-iron displayed in Figure~\ref{fig:cp}. 
The experimental measurement of the heat capacity C$_p$ diverges at the magnetic Curie transition, characteristic of a second-order phase transition \cite{chandler1987introduction}. 
Without a scalable coupled spin-lattice dynamics simulation environment, that properly accounts for thermal expansion and magnetic contribution to the pressure, reproducing the divergence of C$_p$ (and of other thermomechanical properties) at the critical point is not possible. 

Accurate numerical simulations are critical for enabling technological advances, as they shape our fundamental understanding of the underlying solid state physics that dictates material behavior.
Developing high fidelity models however is challenging, because it necessitates capturing physical phenomena that occur across several length and time scales.
This can only be achieved with sufficiently accurate multiscale simulation tools \cite{Hor2012:future, van2020roadmap}, which is the focus of this work. 

Classical molecular dynamics (MD) simulations~\cite{AW1959:studies} provide a useful framework for multiscale modeling by leveraging interatomic potentials (IAPs) to represent the dynamics of atoms on a Born-Oppenheimer potential energy surface (PES) \cite{rapaport2004art}. 
By utilizing massively parallel algorithms \cite{plimpton1995fast} and long time-scale methodologies\cite{voter2002extending}, MD enables bridging \emph{first-principles} with continuum-scale simulations \cite{zepeda2017probing}.

The absorption of machine learning (ML) techniques into the creation of interatomic potentials has lead to classical MD simulations that approach the accuracy of \emph{first-principles} methods.
A large number of these highly accurate ML-IAPs\cite{HBC+2017:universal,Smith2017, ZHW+2018:deep, BPKC2010:gaussian, JNG2014:general, LSB2018:hierarchical,  TST+2015:spectral} have been developed. In general, they are parameterized on training data (configuration energy, atomic forces) from \emph{first-principles} methods like density functional theory (DFT)~\cite{KS1965:selfconsistent} and utilize different flavors of ML model forms to construct the PES.
While they have proven to be useful for large-scale simulations of materials properties \cite{li2020complex, cusentino2020suppression}, further progress in multiscale modeling is hampered by the limitation of ML-IAPs to non-magnetic materials phenomena. 
Even with highly accurate ML-IAPs, state-of-the-art MD simulations cannot reproduce the divergent behavior of C$_p$ near the critical point (Figure~\ref{fig:cp}) because they fail to account for the magnetic degrees of freedom~\cite{dragoni2018achieving}.

Coupling atomic spin dynamics with classical MD has been pioneered by Ma \emph{et al.} \cite{ma2008large,ma2016spilady,ma2020atomistic}. Herein, a classical magnetic spin is assigned to each atom in addition to its position leading to a 6N-dimensional PES (5N if the magnetic spin norms are fixed), instead of the common 3N-dimensional PES in classical MD:
\begin{equation}
    E = \sum_{i=1}^{N} \epsilon\left(\{\bm{r}_{ij},\bm{s}_i\}\right)\,,
    \label{sl_pes}
\end{equation}
where $\bm{r}_{ij}=\bm{r}_{i}-\bm{r}_{j}$ denotes the relative position between atoms i and j, $\bm{s}_i$ the classical spin assigned to atom $i$, and $N$ the number of atoms in the system. 
In most classical spin-lattice calculations, the 6N-dimensional PES is constructed by introducing an atomic spin model on top of a mechanical IAP \cite{ma2008large}. For example, a common approach is to combine a distance-dependent Heisenberg Hamiltonian with an embedded-atom-method (EAM) potential \cite{tranchida2018massively,ma2020atomistic}.  

While these prior approaches recover experimental properties on a qualitative level \cite{dos2020size,zhou2020atomistic}, their combined representation of phononic and magnetic degrees of freedom is not sufficiently consistent for providing quantitative predictions at the level of \emph{first-principles} results. 
More recently, Ma \emph{et al.} developed a magneto-elastic IAP for magnetic iron based on data from \emph{first-principles} calculations \cite{ma2017dynamic}.
However, this remained an isolated attempt as there is no general methodology for generating a magneto-elastic PES in a classical context that enables large-scale spin-lattice dynamics simulations for any magnetic material.

In this work, we overcome this methodological obstacle by providing a data-driven framework for generating magneto-elastic ML-IAPs that (1) provide a consistent representation of both mechanical and magnetic degrees of freedom and (2) achieve near \emph{first-principles} accuracy.
We refer to our new class of IAPs as "magneto-elastic ML-IAPs" as they generate a consistent PES accurately representing the magnetic degrees of freedom and the interplay between magnetic and elastic phenomena.
Our framework couples an atomic spin model (Heisenberg Hamiltonian) with an ML-IAP and provides a unified magneto-elastic PES which yields the correct mechanical forces on the atoms in the MD framework.
The Heisenberg Hamiltonian is parameterized with data from DFT spin-spiral calculations at different degrees of lattice compression. In constructing the ML-IAP, we leverage the flexible and data-driven spectral neighbor analysis potential (SNAP) methodology \cite{TST+2015:spectral} which is trained on a database of magnetic configurations generated using DFT calculations.

We apply our framework to generate a magneto-elastic ML-IAP for the $\alpha$ phase of iron.
We demonstrate that our potential is transferable to an extended area of the phase diagram, corresponding to 
a temperature and pressure range of 0 to 1200~K and 0 to 13~GPa (up to the $\alpha \to \gamma$ and $\alpha \to \epsilon$ transitions, respectively). 
The Curie temperature, which experimentally occurs at approximately 1045~K, lies within this parameter space.
After presenting our training workflow, the "Results" section will probe the "quantum-accuracy" of our magneto-elastic ML-IAP by performing magneto-static comparisons to \emph{first-principles} measurements.
We then stress that our generated magneto-elastic ML-IAP can also be directly used in the LAMMPS package~\cite{plimpton1995fast} to perform magneto-dynamic simulations that take into account both the thermal expansion of the lattice and magnetic pressure due to spin disorder. 
This enables us to maintain a constant ambient pressure throughout all calculations of thermomechanical properties, consistent with conditions prevalent in experiments.
As illustrated in Figure~\ref{fig:cp}, our framework allows us to perform the first pressure-controlled quantitative prediction of the critical behavior across a second-order phase transition within a classical spin-lattice dynamics simulation.

\section*{Results}

\begin{figure}[!ht]
\centering
\includegraphics[width=0.95\columnwidth]{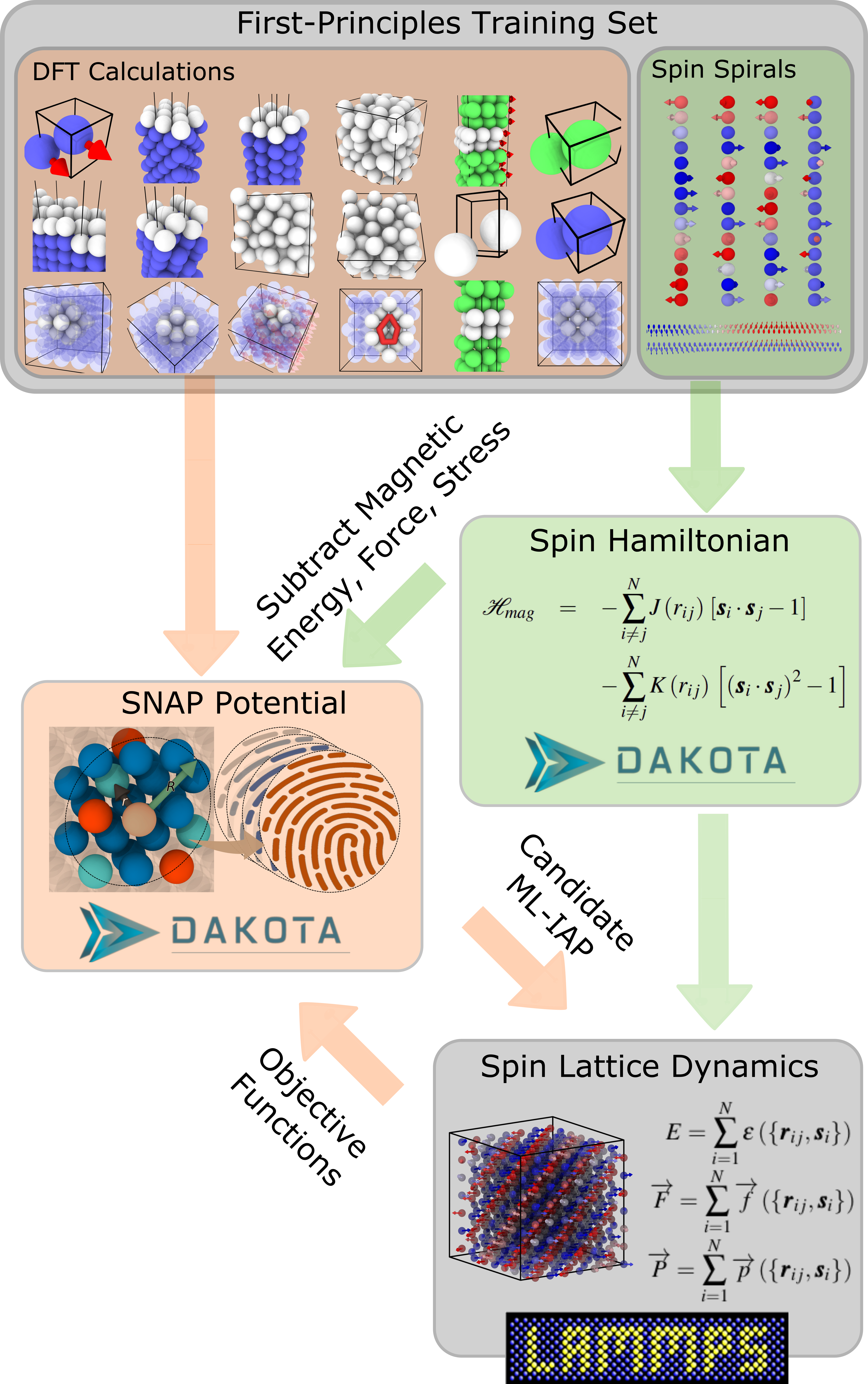}
\caption{Magneto-elastic ML-IAP training workflow. A training set of DFT calculations is partitioned into those that train the SNAP interatomic potential and those that train the spin Hamiltonian, respectively. A \textit{non-magnetic} interatomic potential is fit to configuration energies and atomic forces after the spin Hamiltonian contribution is subtracted and is validated against magneto-elastic properties computed in LAMMPS. Optimization of the spin Hamiltonian and interatomic potential parameters is handled by DAKOTA. 
}
\label{fig:workflow}
\end{figure}

In this section we outline our advancements in magnetic materials modeling.
We first present our training workflow and subsequently assess our results by comparing both static and dynamic properties in $\alpha$-iron against \emph{first-principles} calculations and experiments.

Figure~\ref{fig:workflow} displays our training workflow. Further details to each box in this diagram are presented as a subsection in the "Methods" section. 
All atomic configurations in the training set result from \emph{first-principles} calculations performed with the same DFT setup (same pseudo-potential and energy cutoff, similar k-point densities) as detailed in the "Methods" section.
In contrast to traditional force-matching approaches in the development of classical IAPs, we treat the magnetic and phononic degrees of freedom in the PES in a consistent and unified manner, as indicated by the exchange of information between spin Hamiltonian and SNAP potential parametrization steps.
After parameterizing our atomic spin Hamiltonian by leveraging DFT spin-spiral results, its energy, forces, and stress contributions are subtracted from each atomic configuration in the \emph{first-principles} training set. 
The ML-IAP is then trained to reproduce the \textit{non-magnetic} component of the \emph{first-principles} data. 
Finally, both components of the magneto-elastic PES are recombined to construct a unified magneto-elastic ML-IAP that is consistently trained on \emph{first-principles} data.
Optimization is handled by the DAKOTA software package\cite{eldred2006dakota} in both fitting steps. 
For the SNAP component of the potential, DAKOTA optimizes the radial cutoff of the interaction along with the weights of each training data set (energy and force weights) to generate different candidate potentials. Those candidates potentials are then recombined with the spin Hamiltonian and tested against selected objective functions (mean-absolute errors (MAEs) in lattice constants, cohesive energies, elastic constants, forces and total energies). 
Table~\ref{tab:groups} summarizes the different groups of training data, the optimal weights obtained for each of those groups, and the corresponding energy and force MAEs.
The target values for the objective functions are based on both experimental and DFT data, as outlined in Table \ref{tab:dakota}.
Objective function evaluations are done within LAMMPS~\cite{plimpton1995fast}.

Herein, the critical innovation that enables a leap forward in predictive simulations of magnetic materials is this data-driven workflow. 
Magnetic and phononic contributions to the PES are taken into account explicitly and any miscounting is avoided 
(for example, no double counting of the magnetic energy or contribution to the pressure).
The obtained magneto-elastic ML-IAP can directly be used to run spin-lattice calculations in LAMMPS \cite{plimpton1995fast,thompson2015spectral,tranchida2018massively}.

\subsection*{Magneto-Static Accuracy}\label{ssec:eos_spiral}

\begin{figure*}[ht]
\centering
\includegraphics[width=\textwidth]{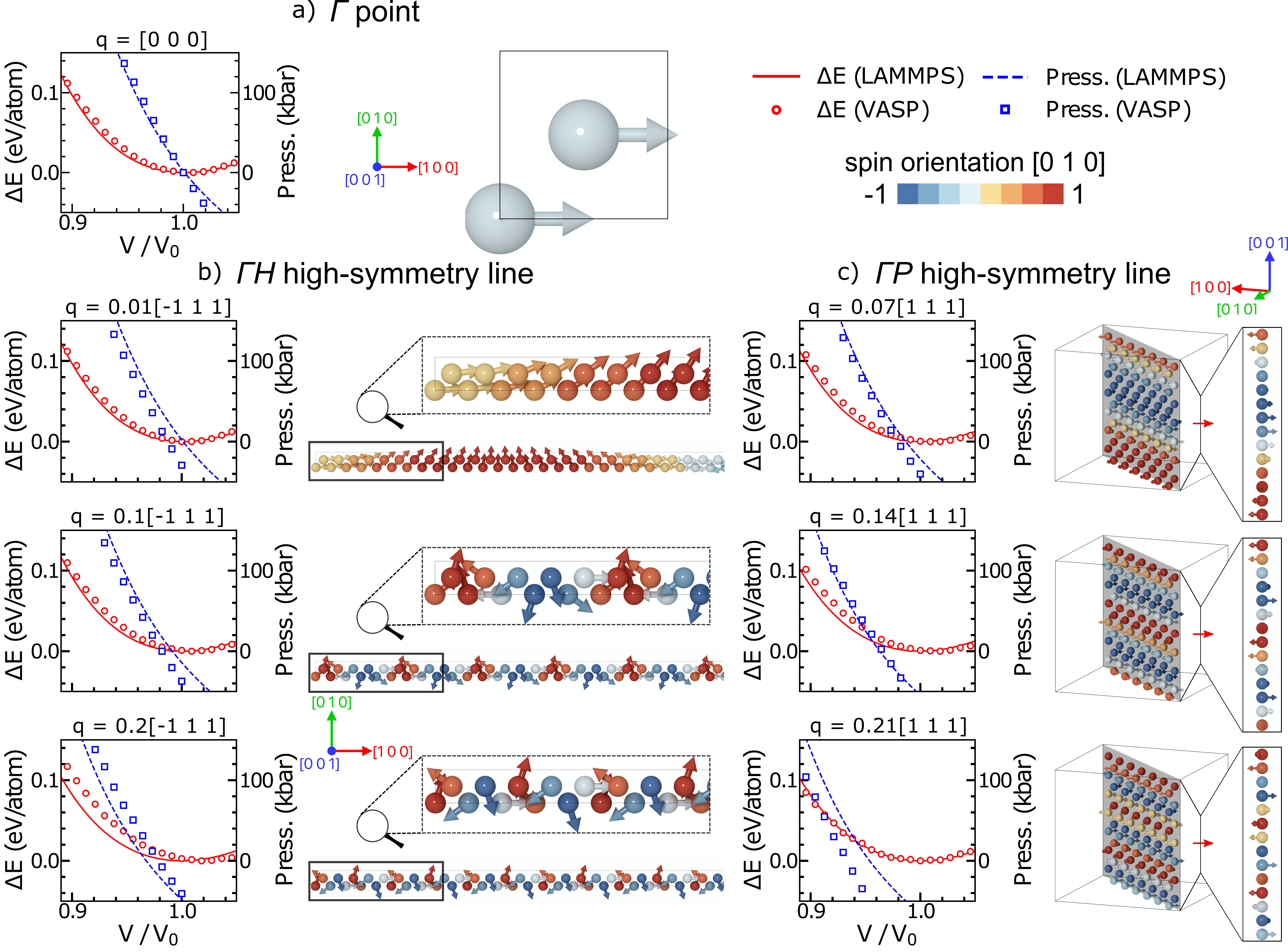}
\caption{Plots of the equation of state data from \emph{first-principles} calculations (VASP computations) and our magneto-elastic ML-IAP (LAMMPS computations) for seven different spin-spirals: a) $\Gamma$ point b) vectors along the $\Gamma H$ high-symmetry line, and c) vectors along the $\Gamma$P high-symmetry line. Visualizations of the corresponding spin-spiral supercells and associated q-vectors are shown to the right of and above each plot, respectively.}
\label{fig:eos_spirals}
\end{figure*}

We first assess the quantitative agreement of our magneto-elastic ML-IAP by comparing with DFT results where magnetic order and elastic deformations are coupled. 
This is done by leveraging a particular subset of spin configurations referred to as spin-spirals, for which the energy and corresponding pressure can be evaluated from both DFT and classical magneto-elastic potential calculations. Details about definition and computation of spin-spirals can be found in the "Methods" section.
Equation-of-state calculations (energy and pressure versus volume) are performed at the $\Gamma$ point (corresponding to the purely ferromagnetic state) and for spin-spirals corresponding to q-vectors along the $\Gamma$H and $\Gamma$P high-symmetry lines.
The calculations at the $\Gamma$ point represent the magnetic ground state and, hence, serve as a point of reference for the spin spiral calculations. 
The geometric orientation of the various computed spin spirals  is visualized in Figure~\ref{fig:eos_spirals}. 
The first set ($q=0.01$ along $\Gamma H$ and $q=0.07$ along $\Gamma P$) represents "long" spirals, close to the $\Gamma$ point, 
the second set ($q=0.1$ along $\Gamma H$ and $q=0.14$ in $\Gamma P$) represents spirals with intermediate periodicity, 
and the last set ($q=0.2$ along $\Gamma H$ and $q=0.21$ along $\Gamma P$) is chosen close to the borders of the magnetic training set (see red demarcation lines in Figure~\ref{fig:fit_spiral} in the "Methods" section). 
The DFT results are obtained by leveraging the generalized Bloch theorem, whereas our classical spin-lattice calculations were performed by generating the corresponding supercells (details given in the "Methods" section). 

Excellent agreement between our classical spin-lattice model and DFT is achieved at the $\Gamma$ point and for the two first q-vectors on each high-symmetry line ($q=0.01$ and $q=0.1$ along $\Gamma H$, $q=0.07$ and $q=0.14$ along $\Gamma P$) in the pressure range relevant for the $\alpha$-phase of iron (up to 13 GPa which corresponds to the $\alpha ~ \to ~ \epsilon$ transition). 
At higher q-vector values, the energy and pressure predictions of our atomic spin-lattice model still agree reasonably well with the DFT calculations.
The observed deviation from the DFT results can be explained by the limitations of our atomic spin-lattice model: as both the pressure and the relative angle between neighboring spins increase, fluctuations of the atomic spin norms become more important. As discussed in the "Methods" and "Discussion" sections, these are not included in the Hamiltonian of our atomic spin-lattice model.

\subsection*{Magneto-Dynamic Accuracy}\label{ssec:application}

\begin{figure*}[!ht]
\centering
\includegraphics[width=\textwidth]{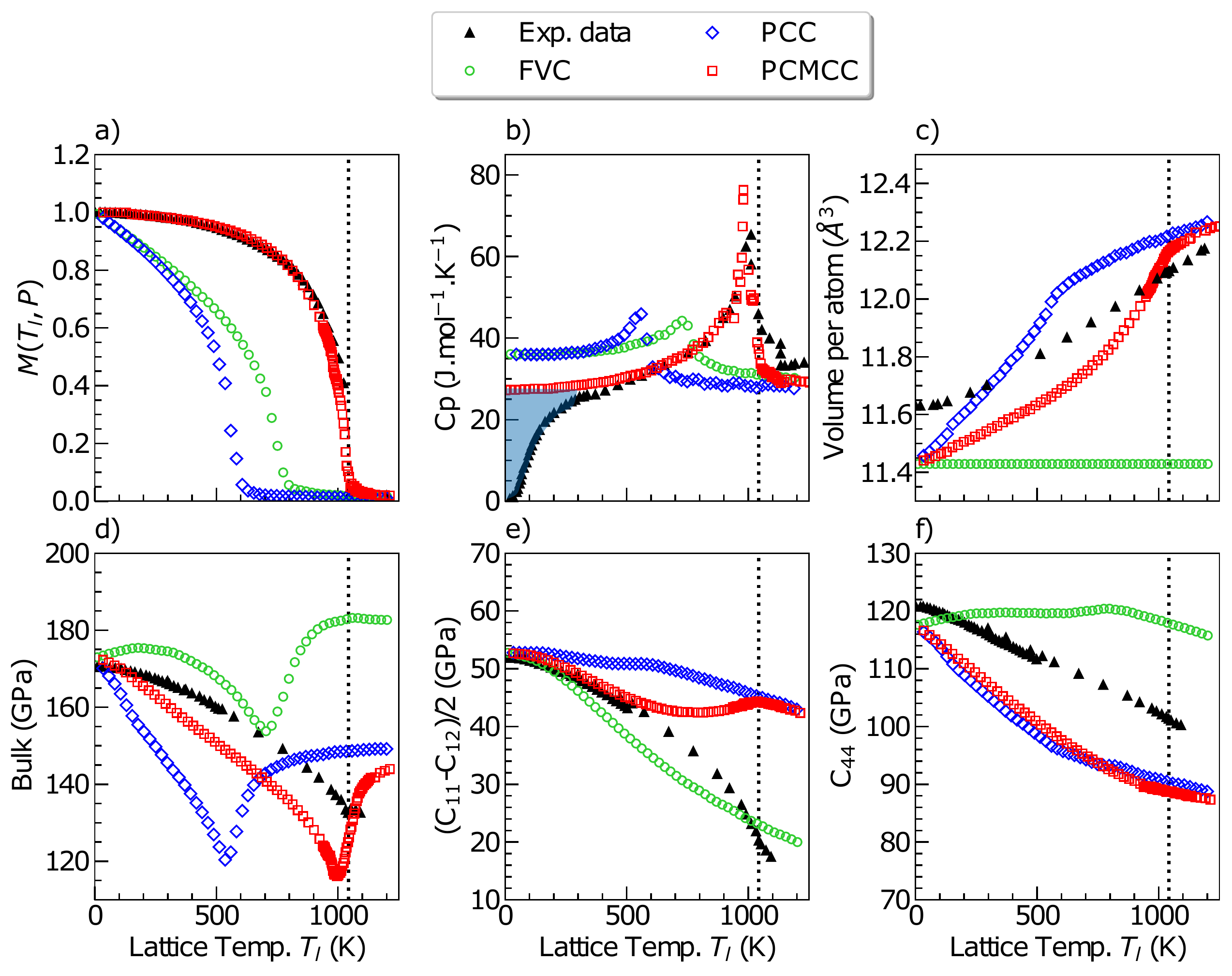}
\caption{Plots a-f show magnetoelastic data obtained with our magneto-elastic ML-IAP. The green (\markerone{}), blue (\markertwo{}), and  red (\markerthree{}) markers indicate the choice of equilibration conditions: 
"fixed-volume conditions" (FVC), "pressure-controlled conditions" (PCC) and "pressure-controlled and magnetization-controlled conditions" (PCMCC), respectively. In all plots, experimental data (extracted from five different references~\cite{crangle1971magnetization,touloukian1970thermophysical,wallace1960specific,seki2005lattice,basinski1955lattice}) is denoted by the filled triangles ($\blacktriangle$), and the dotted black lines (\protect\blackline) represent the experimental Curie temperature. 
The plots in a-b) show magnetization and specific heat comparisons between different ensembles and experiments. 
The light blue region in (b) indicates the low temperature regime $T\lesssim$250~K where quantum effects reduce the experimental heat capacity below the classical Dulong-Petit limiting value of $3R$~\cite{ashcroft2016solid}. 
The data in plot c) illustrates how the lattice expands with temperature. An inherent offset exists between our model (trained to match the DFT data at 0 K) and experimental measurements. 
Plots d-f show (d) bulk modulus, (e) $(c\textsubscript{11} - c\textsubscript{12})/2$ shear constant, and (f) $c\textsubscript{44}$ shear constant for the three aforementioned sets of conditions. 
}
\label{fig:elastic}
\end{figure*}

Turning now to spin-lattice dynamics calculations based on our magneto-elastic ML-IAP (as detailed in the "Methods" section), we assess the quantitative accuracy with respect to experimental measurements of changes in magnetic and thermoelastic properties as the material is heated.
In making this comparison, it is necessary to choose which thermodynamic state variables will be held fixed and which will be allowed to vary with temperature.
Spin-lattice dynamics algorithms have been developed for simulations in a canonical ensemble (CE) which preserves the number of particles, the volume, and the temperature in the system~\cite{ma2020atomistic}.  Our first set of simulation conditions, referred to as "fixed-volume conditions" (FVC), hold the volume fixed while running dynamics in the CE at specified values of the lattice and spin temperatures.
A disadvantage of this choice is that the pressure steadily increases as heat is added to the material, in contradiction to the experimental observations, which are conducted at constant pressure.
To this date, an isobaric spin-lattice algorithm has not been developed (preserving the system's pressure rather than its volume).
However, our methodology as implemented in LAMMPS enables us to compute the magnetic contribution to the pressure. By alternating thermalization (coupled spin-lattice dynamics in a CE) and pressure equilibration (frozen spin configuration in an isobaric ensemble) steps, it is possible to control the pressure of our spin-lattice system. Hence, we refer to calculations performed in this pressure-controlled CE as "pressure-controlled conditions" (PCC).
In both conditions, the temperature of the spin and lattice subsystems is set using two separate Langevin thermostats (one acting on the spins, the other on the lattice) \cite{ma2020atomistic}. 
Finally, this enables us to define a third set of conditions: in addition to controlling the pressure, the spin thermostat can be set to match a given magnetization value (i.e., the experimental magnetization) rather than a temperature. 
We refer to this as "pressure-controlled and magnetization-controlled conditions" (PCMCC).
Figure~\ref{fig:pressure} in the "Methods" section displays the different definitions of the spin temperature and the evolution of the pressure for those three different conditions.

In practice, FVC, PCC and PCMCC only differ in their equilibration conditions (control of pressure and / or magnetization), as each of the corresponding simulations are performed in a canonical ensemble.
We illustrate the predictive capability of our magneto-elastic ML-IAP in $\alpha$-iron for these equilibration conditions in Figure~\ref{fig:elastic}.a-f (FVC : \markerone \hspace{0.5mm}, PCC : \markertwo \hspace{0.5mm},  PCMCC : \markerthree \hspace{0.5mm}).
The agreement of the following magneto-elastic properties with experimental results is assessed: magnetization (Figure~\ref{fig:elastic}.a), heat-capacity C\textsubscript{p} (Figure~\ref{fig:elastic}.b), thermal expansion (cell volume on Figure~\ref{fig:elastic}.c),
bulk modulus (Figure~\ref{fig:elastic}.d), and two shear constants, $(c\textsubscript{11} - c\textsubscript{12})/2$ and $c\textsubscript{44}$ (Figure~\ref{fig:elastic}.e-f).
The "Spin-Lattice Dynamics" subsection of the "Methods" section details the computation of those temperature-dependent elastic constants.

We first work under the FVC (\markerone), keeping a constant volume and equal spin and lattice temperatures (Figure~\ref{fig:elastic}.c and Figure~\ref{fig:pressure}).
At constant volume, our model predicts a Curie temperature of approximately 716K (Figure~\ref{fig:elastic}.a). 
Specific heat calculations shown in Figure~\ref{fig:elastic}.b were performed by computing the derivative of the internal energy, taking both the lattice and magnetic contributions into account. 
The SNAP contribution (lattice only) was first isolated and determined to be a constant value of 26.4 $\rm{J mol}^{-1} \rm{K}^{-1}$, in good agreement with the Dulong-Petit value of $3R$~\cite{ashcroft2016solid}.
The magnetic contribution offsets the total specific heat at low temperature, as the magnetization steadily decreases (thus steadily increasing the magnetic energy).
Also at low temperature, deviation between simulations and experiment (highlighted by the semi-transparent blue region in Figure~\ref{fig:elastic}.b) occurs due to quantum effects which reduce the experimental heat capacity below the $3R$ value.
The FVC heat-capacity is determined at constant volume, although we use the symbol C\textsubscript{p} on the axis label because the enhanced simulations described below are indeed conducted at constant pressure conditions. 
In those constant volume conditions, the pressure evolution with temperature increase is substantial (up to 12 GPa, almost corresponding to the $\alpha \to \epsilon$ transition, as can be seen on Figure~\ref{fig:pressure}), which has a strong impact on the underlying elastic properties. 
Interestingly, at the Curie temperature (here 716K), the increasing pressure exhibits an inflection point, confirming the importance of spin fluctuations on the thermoelastic properties.
The temperature dependence of three elastic constants is shown in Figure~\ref{fig:elastic}.d-f. 
For the bulk modulus, FVC does not agree well with experimental data, especially at higher temperatures. The FVC results tend to overestimate the stiffness, which most likely arises from the build-up of thermal stresses in the material. Under these conditions a nearly temperature-invariant c\textsubscript{44} response is predicted, which is in strong contrast to trends in experiment. Despite these shortcomings, the FVC calculations actually match the experimental data for shear constant $(c\textsubscript{11} - c\textsubscript{12})/2$ relatively well throughout the entire temperature range. 
In general, the fixed volume assumption made under FVC fails to account for thermal expansion, leading to incorrect elastic predictions. 

We correct this shortcoming of the model by working under PCC (\markertwo) which allows for thermal expansion.
As can be seen on Figure~\ref{fig:elastic}.c, the cell volumes are relaxed at each finite temperature, until the pressure in the system drops to approximately 0 GPa.
As shown in Figure~\ref{fig:elastic}.a, the thermal expansion incorrectly moves the onset of Curie transition to approximately 536K. 
As the average interatomic distance increases, the strength of the exchange interaction is lowered, thus decreasing the transition temperature.
The computed heat-capacity (Figure~\ref{fig:elastic}.b) now corresponds to the derivative of the free energy, and to an actual C\textsubscript{p} measurement. 
However, as in the FVC, the low agreement between the experimental and computed magnetization evolution leads to an offset in the initial C\textsubscript{p} and does not match the Dulong-Petit value at low temperature.
The PCC fares better in reproducing the experimental bulk modulus up to the Curie transition (no hardening observed).
PCC also does better in terms of the shear constant c\textsubscript{44}, as it is able to reproduce the thermal softening seen in experiments. 
However, for shear constant $(c\textsubscript{11} - c\textsubscript{12})/2$, PCC underestimates the extent of the thermal softening. 
Overall, PCC does better than FVC in terms of elastic properties, but deviates more in terms of magnetic predictions compared to experiment. 
By shifting the Curie transition towards lower temperatures, it reduces the range of validity of our elastic calculations. 

In order to improve the magnetic predictions of $\alpha$-iron, we finally consider the PCMCC scheme (\markerthree).
In addition to allowing for thermal expansion similarly to the PCC, we also set the spin thermostat temperature in order to reproduce the experimental magnetization.  Below the Curie transition, the spin temperature increases more slowly than the lattice temperature, while above the Curie transition, it increases at the same rate as the lattice temperature (see Figure~\ref{fig:pressure} in the "Methods section").
Figure~\ref{fig:elastic}.a shows that the obtained magnetization under PCMCC closely matches that of experiment.
Most prominently, the resulting C\textsubscript{p} agrees well with experiments (Figure~\ref{fig:elastic}.b). The Dulong-Petit value is recovered at low temperature, and the C\textsubscript{p} discontinuity at the Curie transition is well captured.
The thermal expansion trend is also in much better agreement with experiments, with very comparable slopes between approximately 200 and 750K (Figure~\ref{fig:elastic}.c).
Up to approximately 600 K, PCMCC agree very well with the experimental values for $(c\textsubscript{11} - c\textsubscript{12})/2$ (Figure~\ref{fig:elastic}.e) but at 800-1000K a slight hardening is observed, which contradicts experimental data. 
For the bulk modulus, PCMCC correctly predicts the nearly linear trend up to the Curie temperature. 

We note that in all three sets of conditions, a rapid increase of about 25-30 GPa in the bulk modulus is observed as we move across the critical point. 
This jump was found to be strongly impacted by the underlying mechanical potential. The prediction accuracy could possibly be improved by including additional, finite-temperature objective functions in the fitting procedure. 
The PCMCC prediction of the shear constant c\textsubscript{44} closely matches the PCC data. This tends to indicate that this shear constant c\textsubscript{44} is not impacted significantly by the spin dynamics.
For both pressure controlled conditions (PCC and PCMCC) the maximum deviation from experiments occurs near 700K and is approximately 14\%.

\section*{Discussion}

We presented a data-driven framework for automated generation of magneto-elastic ML-IAPs which enable large-scale spin-lattice dynamics simulations for any magnetic material in LAMMPS.
This framework was demonstrated by generating a robust magneto-elastic ML-IAP for $\alpha$-iron.
First we investigated the magneto-static accuracy (energy and pressure) with respect to equivalent \emph{first-principles} calculations. It was demonstrated that the generated magneto-elastic ML-IAP (which represents the corresponding 5-N dimensional PES) is in close agreement with \emph{first-principles} magneto-elastic calculations. This was achieved by properly partitioning the PES into magnetic and mechanical degrees of freedom.
Subsequently, we investigated the magneto-dynamic accuracy by comparing predicted finite temperature magneto-elastic properties (magnetization, heat-capacity, thermal-expansion, bulk modulus, and shear constants) across the ferromagnetic-paramagnetic phase transition from spin-lattice dynamics simulations against data from experiments. 
In the course of this, we analyzed the choice of simulation conditions (control of pressure and magnetization) and highlighted the importance of thermal and magnetic pressure contributions.
This is an important advance over traditional classical magnetization dynamics methods, where contributions from thermal expansion or spin pressure due to disorder are negated.  
We demonstrated that spin-lattice dynamics simulations of controlled pressure and constrained magnetization yields qualitative agreement with the measured magneto-elastic properties.

Our framework enables predictions of critical properties across the second-order phase transition within classical spin-lattice dynamics simulations, such as the divergent behavior of the heat capacity around the Curie temperature (Figure~\ref{fig:cp} and \ref{fig:elastic}.b).
We provide a more comprehensive perspective on our results by comparing them within the context of other first-principles and classical methods.
At low temperature, \emph{first-principles} methods can capture the electronic component of the heat-capacity, up to the Dulong-Petit value \cite{dragoni2015thermoelastic,ashcroft2016solid} (the difference with our model is highlighted by the blue area on Figure~\ref{fig:elastic}.b).
However, computing C$_p$ across the Curie transition requires a dynamic treatment of large spin-ensembles whose calculation  is computationally expensive in terms of \emph{first-principles} methods.
Classical IAPs do not explicitly treat magnetic degrees of freedom and, thus, cannot reproduce the effects of this magnetic phase-transition \cite{dragoni2018vibrational}.
An empirical model which is based on \emph{first-principles} calculations and accounts for electronic, phononic and magnetic degrees of freedom gave excellent agreement with the experimental C$_p$ curve of $\alpha$-iron up to the Curie temperature \cite{kormann2008free}. 
However, this model does not extend above the Curie temperature, does not account for the pressure generated by the corresponding spin configurations, and cannot be easily extended to other thermomechanical properties.
Thus, for a range of temperature from about 250 K to 1200 K, our model provides with a set of very good predictions, obtained for the computational cost of classical MD calculations only.

We conclude the discussion of our results by pointing out limitations of the present method and future prospects.
First, note that the agreement to the experimental Curie transition (T$_c \approx 716$K in a fixed volume calculation) could have been adjusted by parameterizing the spin potential on a smaller range of the high-symmetry lines (see Figure~\ref{fig:fit_spiral}), or by adding an objective function aimed at matching the experimental value in the spin-potential fitting procedure.
However, this additional constraint would have worsened the agreement of our model with the DFT energy and pressure results (as displayed on Figure \ref{fig:eos_spirals}) and would contradict the overall objective of this work.

For temperatures below approximately 250 K, our classical framework cannot access the quantized free energy, and is thus enable to accurately reproduce the trends of all the quantities being its derivatives (C$_p$, elastic constants, ...).
This is reducing the agreement versus experiments of the magneto-dynamic accuracy measurements displayed on Figure~\ref{fig:elastic} at low temperature, and can be seen as a limitation of our classical approach \cite{arakawa2020quantum}. 

Another limitation of our work lies in the simplicity of the spin Hamiltonian model used. 
Extended spin Hamiltonians, such as spin-cluster expansions, might be a promising route to improving the accuracy of the magnetic component of the PES by both accounting for the fluctuation of the magnetic moment magnitudes and many-body spin interactions \cite{drautz2004spin,drautz2020atomic}. 
A straightforward extension of this work could combine recently developed extended spin Hamiltonians with \emph{first-principles} studies, and apply our formalism to extend our $\alpha$-iron magneto-elastic ML-IAP to account for defect configurations~\cite{marinica2012irradiation,chapman2020effect}, Cr clustering \cite{chapman2019dynamics,klaver2006magnetism}, and magneto-structural phase-transitions~\cite{surh2016magnetostructural,kalantar2005direct}.

Enhanced magnetic thermostats have also been proposed in order to better match the experimental magnetic transition versus temperature~\cite{woo2015quantum,bergqvist2018realistic}.
Such thermostats could be implemented in LAMMPS and used to replace the magnetization-controlled conditions defined in the "Results" section. 
This could extend the range validity of our framework to areas of phase-diagrams where the magnetization distribution is not well measured (for example in the $\epsilon$ phase of iron).

A recent study added a magnetic contribution to the set of descriptors used in a moment-tensor ML-IAP~\cite{novikov2020machine}. Although this approach does not explicitly simulate the magnetization dynamics (and its effects on thermomechanical properties), the authors demonstrated remarkable improvement in terms of error convergence.
At this stage of our work, we believe improving the modeling of the magnetic component of the PES remains our first priority (and thus implementing and fitting improved spin Hamiltonians, as discussed above). 
However, depending on the success of this first effort, this complementary approach could be leveraged to improve the quantum-accuracy of our magneto-elastic ML-IAPs.

In summary, we have presented a new computational framework for near quantum-accuracy simulations of magneto-elastic materials properties.
By leveraging the flexibility of ML-IAPs, our data-driven workflow enables to model the interplay between magnetic and phononic dynamics for a large class of magnetic materials.
Furthermore, our straightforward connection to the LAMMPS package makes it possible to perform large-scale quantitative magneto-elastic predictions over controlled pressure and temperature spaces, hitherto study unexplored magneto-dynamics properties of materials.

\section*{Methods}

\subsection*{Density functional theory calculations}\label{ssec:dft}

Parameterizing both the ML-IAP and the magnetic Heisenberg Hamiltonian relies on data computed using spin-dependent DFT calculations. They were performed using VASP~\cite{KRESSE199615,PhysRevB.59.1758}.
In all calculations the PBE~\cite{perdew1996generalized} exchange-correlation functional was employed. 
We used PAW pseudopotentials~\cite{blochl1994projector} with 8 valence electrons and a core radius of $r_c=2.3$~a$_B$. 
The plane wave cutoff was set to $320$~eV and the convergence in each self-consistency cycle was set to $10^{-8}$. 
The Fermi-Dirac smearing scheme with a width of 0.026 eV was used. 
The Brillouin zone was sampled on a $10\times 10\times 10$ grid of \emph{k}-points.
The number of bands used was 224 per atom.


\subsection*{Spin-Spiral Calculations}\label{ssec:spirals}

Spin-spirals define a subset of non-collinear magnetic states. 
In this work, we leverage spin-spirals as a convenient tool to perform one-to-one comparisons between \emph{first-principles} and classical magneto-elastic calculations. They can be defined as follows:
\begin{equation}
  \bm{s}_{j} = \sin\theta\cos(\bm{q}\cdot\bm{R}_{0j})\bm{\hat{x}}+
               \sin\theta\sin(\bm{q}\cdot\bm{R}_{0j})\bm{\hat{y}}+
               \cos\theta\bm{\hat{z}}\,,
  \label{spiral}
\end{equation}
where $\bm{q}$ is the spin-spiral vector, $\bm{R}_{0j}$ is the position of atom $j$ relative to a central atom $0$, $\bm{s}_{j}$ is the spin on atom $j$, and $\theta$ is a constant angle between the spins and the spin-spiral vector (often referred to as "cone angle") \cite{zimmermann2019comparison}.  
$\bm{\hat{x}}$,  $\bm{\hat{y}}$, and $\bm{\hat{z}}$ are the unit vectors along $[100]$, $[010]$, and $[001]$, respectively. Our calculations are restricted to $\theta=\pi/2$, corresponding to flat spin-spirals in the (001) plane.

\emph{First-principles} calculations of the per-atom energy and the pressure corresponding to spin-spiral states are performed using DFT by leveraging the frozen-magnon approach \cite{halilov1997magnon,kurz2004ab} and the generalized Bloch theorem \cite{sandratskii1998noncollinear} as implemented in VASP \cite{marsman2002broken}.
We consider a primitive cell of one atom.
A $10\times 10\times 10$ k-point grid, an energy cutoff of 320 eV, and 224 bands proved sufficient to reach the level of accuracy expected in our model (as can be seen in Figure~\ref{fig:fit_spiral}).

Classical calculations are performed by using Eq.~(\ref{spiral}) to generate supercells accommodating the spin-spirals corresponding to the $\bm{q}$-vectors used in the DFT calculations.
Based on a given supercell and a spin Hamiltonian, the per-atom energy and pressure are computed using the SPIN package of LAMMPS \cite{plimpton1995fast,tranchida2018massively}. 

\subsection*{\label{ssec:spin} Spin Hamiltonian}

A spin Hamiltonian is used to model the energy, mechanical forces, and pressure contributions of magnetic configurations.
Rosengaard and Johansson~\cite{rosengaard1997finite}  and Szilva~\emph{et al.}~\cite{szilva2013interatomic} showed that adding a biquadratic term to the classical Heisenberg Hamiltonian improves the accuracy of magnetic excitations in 3-d transition ferromagnets.  
We adopted their Hamiltonian form:
\begin{eqnarray}
\mathcal{H}_{mag} &=& -\sum_{i\neq j}^{N} {J} \left(r_{ij} \right)\,
                      \left[\bm{s}_{i}\cdot \bm{s}_{j} -1 \right] \nonumber \\
                  &~& -\sum_{i\neq j}^{N} {K} \left(r_{ij} \right)\, 
                  \left[\left(\bm{s}_{i}\cdot \bm{s}_{j}\right)^2 -1 \right]\,,
                  \label{s_hamiltonian}
\end{eqnarray}
where $\bm{s}_{i}$ and $ \bm{s}_{j}$ are classical atomic spins of unit length located on atoms $i$ and $j$, ${J} \left(r_{ij} \right)$ and ${K} \left(r_{ij} \right)$ (in eV) are magnetic exchange functions, and $r_{ij}$ is the interatomic distance between magnetic atoms $i$ and $j$.
The two terms in Eq.~\ref{s_hamiltonian} are offset by subtracting the spin ground state (corresponding to a purely ferromagnetic situation), as detailed in Ma \emph{et al.} \cite{ma2008large}. 
Although this offset of the exchange energy does not affect the precession dynamics of the spins, it allows to offset the corresponding mechanical forces. Without this additional term, the magnetic contribution to the forces and the pressure are not zero at the energy ground state.
For the exchange interaction terms ${J}\left(r_{ij} \right)\ $ and ${K}\left(r_{ij} \right)$, the interatomic dependence is taken into account through the following function based on an approximation of the Bethe-Slater curve \cite{kaneyoshi1992introduction, yosida1996theory}:
\begin{equation}
  f\left(r \right) = 4\alpha
  \left(\frac{r}{\delta}\right)^2
  \left(1-\gamma\left(\frac{r}{\delta}\right)^2 \right)
  \exp\left(-\left(\frac{r}{\delta}\right)^2 \right)
  \Theta\left(R_c - r\right)\,,
  \label{J_dep}
\end{equation}
where $\alpha$ denotes the interaction energy, $\delta$ the interaction decay length, $\gamma$ a dimensionless curvature parameter, r a distance, and $\Theta\left(R_c - r\right)$ a Heaviside step function for the radial cutoff $R_c$. 
This assumes that the interaction decays rapidly with the interatomic distance, consistent with former calculations \cite{pajda2001ab, szilva2013interatomic}. 
We set $R_c=~$5\AA ~to include five neighbor shells, as Pajda \emph{et al.}~\cite{pajda2001ab} showed that the exchange interaction decays slower along the $[111]$ direction in $\alpha$-iron.

Using Eq.~(\ref{s_hamiltonian}) and leveraging the generalized spin-lattice Poisson bracket as defined by Yang \emph{et al.}~\cite{yang1980generalizations}, the magnetic precession vectors ($\bm{\omega}_{i}$), mechanical forces ($\bm{F}_{i}$), and their corresponding virial components ($W\left(\bm{r}^N \right)$) are derived:
\begin{eqnarray}
\bm{\omega}_{i} &=& \frac{1}{\hbar}\sum_{j}^{N_i} {J} \left(r_{ij} \right)\,
                      \bm{s}_{j} + K\left(r_{ij} \right) 
                  \left(\bm{s}_{i}\cdot \bm{s}_{j}\right) \bm{s}_{j}\,,                 \label{m_prec} \\
\bm{F}_{i} &=& \sum_{j}^{N_i} \frac{d{J} \left(r_{ij} \right)}{d_{rij}}\,
                \left[\bm{s}_{i}\cdot \bm{s}_{j} -1 \right]\bm{e}_{ij} \nonumber \\
            &~& +\frac{d{K} \left(r_{ij} \right)}{d_{rij}}\,
                \left[\left(\bm{s}_{i}\cdot \bm{s}_{j}\right)^2 -1 \right]\bm{e}_{ij}\,, 
                \label{m_force} \\
W\left(\bm{r}^N \right) &=& \sum_{i=1}^{N} \bm{r}_i\cdot\bm{F}_{i}\,, 
\end{eqnarray}
where $\bm{r}^N$ denotes a $3N$ size vector of all atomic positions and $\bm{r}_i$ the position vector of atom $i$. We note that the virial components enable computing the spin contribution to the pressure. 

\begin{figure}[!ht]
\centering
\includegraphics[width=0.95\columnwidth]{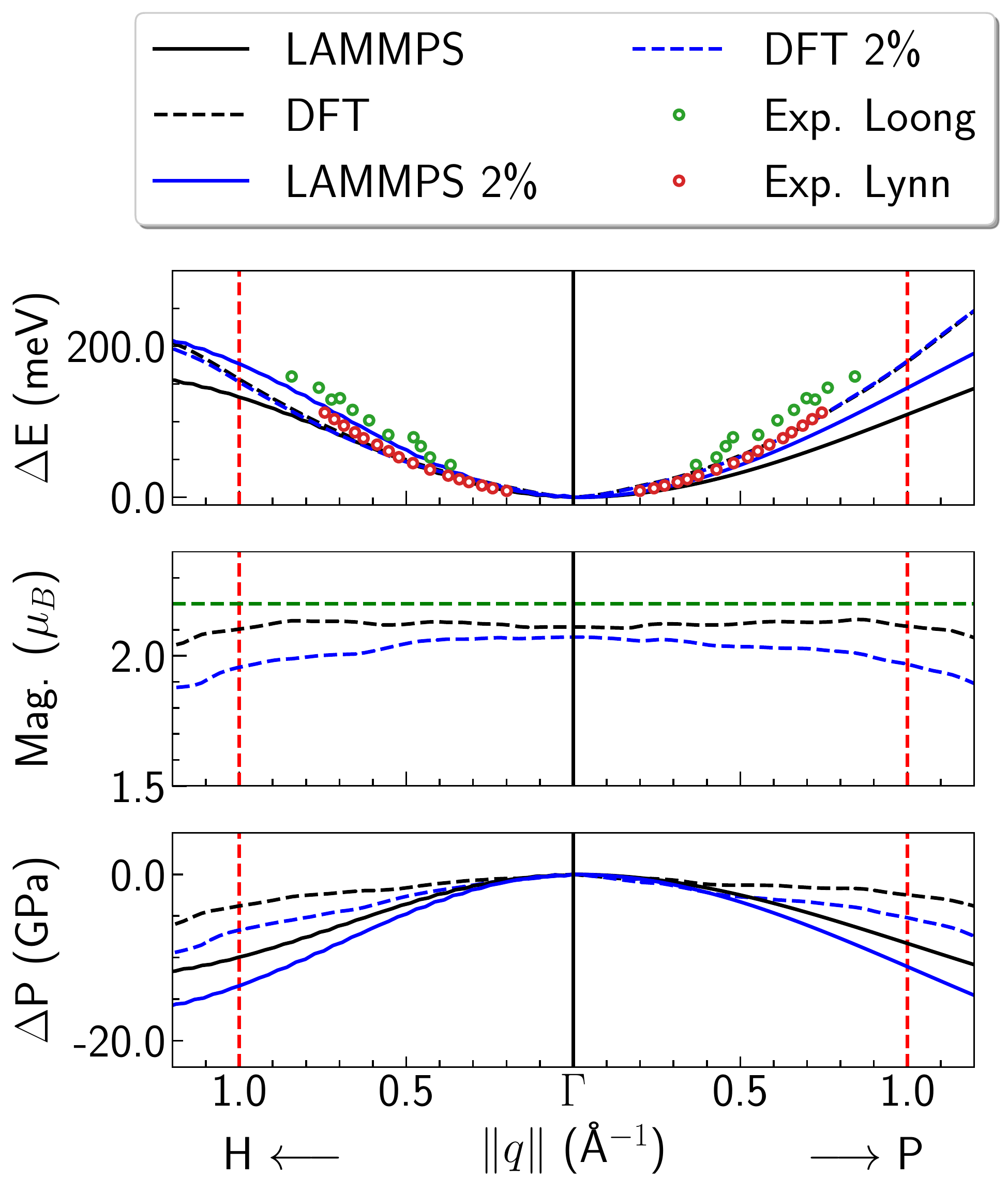}
\caption{
Comparison of spin-spiral results along sections of the $\Gamma$H and $\Gamma$P high-symmetry lines. 
The upper plot displays the per-atom energy, the middle one the atomic moment fluctuations (in Bohr magneton per atom), and on the bottom the evolution of the pressure.
The energy and pressure fluctuations are plotted with respect to the magnetic ground state at the $\Gamma$ point.
The green and red dots represent experimental measurements obtained by Loong \emph{et al.}~\cite{loong1984neutron} and Lynn~\cite{lynn1975temperature}.
In all three plots, the dashed lines correspond to the DFT results, and the continuous lines to our classical model results, whereas the line color  (black or blue) corresponds to the lattice compression (0 or 2\%, respectively). 
In the middle plot, the green dashed horizontal line represent the experimental equilibrium value (2.2 $\mu_B$ per atom), which is the constant value chosen in our model.
In all three plots, the red vertical dashed lines are delimiting the $\bm{q}$-vectors on which our spin Hamiltonian was parametrized.
}
\label{fig:fit_spiral}
\end{figure}

The spin Hamiltonian is used to reproduce spin-spiral energy and pressure reference results obtained from DFT.
They are sampled along two high-symmetry lines, $\Gamma$H and $\Gamma$P, and for two different lattice constant values (corresponding to the equilibrium bulk value and to a lattice compression of 2\%). 
This allows us to encapsulate in the model the influence of lattice compression on the spin stiffness and the Curie temperature, which was experimentally and theoretically predicted to be small~\cite{leger1972pressure,moran2003ab,kormann2009pressure}.
Figure \ref{fig:fit_spiral} displays the excellent agreement obtained between our \emph{first-principles} spin-spiral energies and experimental measurements.

Our current spin Hamiltonian does not account for fluctuations of the magnetic moment magnitudes, i.e. the norm of atomic spins remains constant in our calculations.
As can be seen in Figure~\ref{fig:fit_spiral}, this is not the case for our DFT results, as those fluctuations can become important when departing from the $\Gamma$ point. 
We thus decided to parameterize our model only on spin-spirals corresponding to $\bm{q}$-vectors for which the spin norm deviates from the ferromagnetic value ($\approx 2.2~\mu_B$/atom at the $\rm{\Gamma}$ point) by less than 5\%. The red dashed lines in Figure~\ref{fig:fit_spiral} delimit this $\bm{q}$-vector range. 

Finally, we used the single objective genetic algorithm within the DAKOTA software package \cite{eldred2006dakota} to optimize the six coefficients of ${J}\left(r_{ij} \right)$ and ${K}\left(r_{ij}\right)$ in order to obtain the best possible agreement between our reference DFT spin-spiral energy and pressure results and our spin model.
Figure~\ref{fig:fit_spiral} displays the obtained result. 
As can be seen in Figure~\ref{fig:elastic}, for a fixed-volume calculation, our spin Hamiltonian predicts a Curie temperature of 716K. 
Note that a better match of the DFT spin-spiral energies would yield a larger spin-stiffness, and thus a better agreement for the Curie temperature. However, this would worsen the pressure agreement. 


Spin-orbit coupling effects were included by accounting for an iron-type cubic anisotropy~\cite{skomski2008simple}:
\begin{equation}
\begin{split}
  H_{cubic} &= -\sum_{i=1}^{N} K_1 \bigg[(\bm{s}_{i}\cdot \bm{\hat{x}})^2 (\bm{s}_{i}\cdot \bm{\hat{y}})^2 + (\bm{s}_{i}\cdot \bm{\hat{y}})^2 (\bm{s}_{i}\cdot \bm{\hat{z}})^2 + ...\\ 
& (\bm{s}_{i}\cdot \bm{\hat{x}})^2 (\bm{s}_{i}\cdot \bm{\hat{z}})^2 \bigg] + K_2^{(c)} (\bm{s}_{i}\cdot \bm{\hat{x}})^2 (\bm{s}_{i}\cdot \bm{\hat{y}})^2 (\bm{s}_{i}\cdot \bm{\hat{z}})^2\ ,
\label{cubic_hamiltonian}
\end{split}
\end{equation}
with $K_1 = 0.001$ eV and $K_2^{(c)} = 0.0005$ eV the intensity coefficients corresponding to $\alpha$-iron.
The cubic anisotropy was only included to run calculations, but ignored in the fitting procedure as its intensity is below the range of accuracy of our ML-IAP. 

In all our classical spin-lattice dynamics calculations, our system size remained small compared to the typical magnetic domain-wall width in iron~\cite{skomski2008simple}. Thus, long-range dipole-dipole interactions could safely be neglected.

\subsection*{\label{ssec:snap} SNAP Potential}

For this work, an interatomic potential for iron was developed that is specifically parameterized for use in coupled spin and molecular dynamics simulations.
Training data for a Spectral Neighborhood Analysis Potential (SNAP)\cite{thompson2015spectral, wood2018extending, zuo2020performance} was collected to constrain the fit to the pressure and temperature phase space of $<20$GPa and $<2000$K. 
The set of non-colinear, spin-polarized VASP calculations includes $\alpha$- (BCC), $\epsilon$- (HCP) and liquid-iron, Table ~\ref{tab:groups} displays the quantity of each training type and target properties that are captured therein. 
Optimization of a SNAP potential necessitates that the generated training database be broken into these \textit{groups} (rows in Table ~\ref{tab:groups}) such that the weighted linear regression can (de-)emphasize different parts in search of a global minima in objective function errors.
Each training group is assigned a unique weight for its' associated energies and atomic forces for each candidate potential, optimization of these weights is controlled by DAKOTA. 
Regression is carried out using singular value decomposition with a squared loss function (L2 norm). 
In order to avoid double counting, and properly simulate the magnetic properties of iron in classical MD, we have adapted the SNAP fitting protocol\cite{thompson2015spectral} to isolate the non-magnetic energy and forces from the generated training data. 
To do so, the fitted biquadratic spin Hamiltonian is evaluated for every atom in the training set, and its' contribution to the total energy and per-atom forces is subtracted.
This is akin to previous uses of an ion core repulsion\cite{wood2019data} or electrostatic interaction term\cite{deng2019electrostatic} as a \textit{reference potential} while fitting SNAP models.

\input{table_db}

Optimization of the SNAP potential was achieved using a single objective genetic algorithm within the DAKOTA software package\cite{eldred2006dakota}. 
Radial cutoff distance, training group weights and number of bispectrum descriptors were varied to minimize a set of objective functions, as percent error to available DFT or experimental\cite{adams2006elastic} data, that encapsulate the desired mechanical properties of Fe. 
These objective functions specific to $\alpha$-iron are listed in Table \ref{tab:dakota}, and the RMSE energy and force regression errors are included in optimization as well. 
In all objectives, our linear SNAP model with 31 bispectrum descriptors achieves accuracy in all mechanical properties within a few percent of experiment/DFT. 
Additionally, lattice constants and cohesive energies of $\gamma$- (FCC) and $\epsilon$-iron (HCP) phases were fit, but given far less priority with respect to the $\alpha$-iron mechanical properties resulting in $\sim 6-7\%$ errors with respect to DFT. 
Importantly, each of the objective functions were evaluated including the magnetic spin contributions to avoid unforeseen changes in property predictions.
A full breakdown of the optimal training group weights and mean absolute energy/force errors are given in Table ~\ref{tab:groups}.
Group weights listed have been adjusted by the number of configurations or forces they are applied to, therefore allowing for larger group weights to be (cautiously) interpreted more \textit{valuable} at meeting the set of targeted objective functions. 
This optimized Fe-SNAP interatomic potential is contained as Supplemental Material along with LAMMPS input scripts used in the following section.

\input{table_dakota.tex}

\subsection*{Spin-Lattice Dynamics}\label{ssec:sl}

Calculations are performed following the spin-lattice dynamics approach as implemented in the SPIN package of LAMMPS \cite{plimpton1995fast,tranchida2018massively}, and set by the spin-lattice Hamiltonian below:

\begin{equation}
\begin{split}
  \mathcal{H}_{sl}(\bm{r},\bm{p},\bm{s}) &= \mathcal{H}_{mag}(\bm{r},\bm{s}) + \sum_{i=1}^{N} \frac{{\lvert \bm{p} \rvert ^2}}{2m_i} + \sum_{i,j=1}^{N} V_{SNAP}(r_{ij})
\label{SL_Hamiltonian}
\end{split}
\end{equation}

where $\mathcal{H}_{mag}$ is the spin Hamiltonian defined by the combination of Eq.~(\ref{s_hamiltonian}) and Eq.~(\ref{cubic_hamiltonian}). The term $V_{SNAP}(r_{ij})$ is our SNAP ML-IAP. The second term on the right in Eq.~(\ref{SL_Hamiltonian}), represents the kinetic energy, where the particle momentum is given as $\bm{p}$ and the mass of particle $\it{i}$ is $m_{i}$. Based on this spin-lattice Hamiltonian and leveraging the generalized spin-lattice Poisson bracket as defined by Yang \emph{et al.}~\cite{yang1980generalizations}, the equations of motion can be defined as:

\begin{equation}
\begin{split}
 \hspace{0.8mm} \frac{d\bm{r}_{i}}{dt} = \frac{\bm{p}_{i}}{m_i}
\label{SL_EOM_1}
\end{split}
\end{equation}

\begin{equation}
\begin{split}
  \frac{d\bm{p}_{i}}{dt} &= \sum_{j,i \neq j}^{N} \bigg[ - \frac{dV_{SNAP}(r_{ij})}{dr_{ij}} + \frac{dJ(r_{ij})}{dr_{ij}}(\bm{s}_{i} \cdot \bm{s}_{j}) + ... \\ &\frac{dK(r_{ij})}{dr_{ij}}(\bm{s}_{i} \cdot \bm{s}_{j})^2 \bigg] \bm{e}_{ij} - \frac{\gamma_{L}}{m_i}\bm{p}_{i} + \bm{f}(t)
\label{SL_EOM_2}
\end{split}
\end{equation}

\begin{equation}
\begin{split}
\hspace{0.4mm}  \frac{d\bm{s}_{i}}{dt} &= \frac{1}{(1+\lambda^2)}\bigg[(\bm{\omega}_{i}+\bm{\eta}(t)) \times \bm{s}_{i} + ...\\ &\lambda\bm{s}_{i} \times (\bm{\omega}_{i} \times \bm{s}_{i})\bigg]
\label{SL_EOM_3}
\end{split}
\end{equation}

Particle positions are advanced according to Eq.~(\ref{SL_EOM_1}). The derivative of the momentum, given in Eq.~(\ref{SL_EOM_2}), is dependent not only on the mechanical potential but the magnetic exchange functions as well. Here $\gamma_L$ is the Langevin damping constant for the lattice and $\bm{f}$ is a fluctuating force following Gaussian statistics given below\cite{tranchida2018massively}.

\begin{eqnarray}
  \langle \bm{f}(t) \rangle &=& 0 
\label{fd_lat_1}\\
  \langle f_{\alpha}(t) f_{\beta} (t') \rangle &=& 2 k_B T_l \gamma_L \delta_{\alpha \beta} \delta (t - t') 
\label{fd_lat_2}
\end{eqnarray}

The fluctuating force $\bm{f}$ is coupled to $\gamma_L$ via the fluctuation dissipation theorem as shown in Eq.~(\ref{fd_lat_2}). Here $k_B$ is the Boltzmann constant, $T_l$ is the lattice temperature, and $\alpha$ and $\beta$ are coordinates. Shown in Eq.~(\ref{SL_EOM_3}) is the stochastic Landau-Lifshitz-Gilbert equation which describes the precessional motion of spins under the influence of thermal noise. In Eq.~(\ref{SL_EOM_3}), $\lambda$ is the transverse damping constant and $\bm{\omega}_i$ is a spin force analogue as shown in Eq.~(\ref{m_prec}). The variable $\bm{\eta}(t)$ is a random vector whose components are drawn from a Gaussian probability distribution given below:

\begin{eqnarray}
  \langle \bm{\eta}(t) \rangle &=& 0
\label{fd_spin_1}\\
  \langle \eta_{\alpha}(t) \eta_{\beta} (t') \rangle &=& D_{S} \delta_{\alpha \beta} \delta (t - t')
\label{fd_spin_2}
\end{eqnarray}

where the amplitude of the noise $D_{S}$ can be related to the temperature of the external spin bath $T_s$ according to $D_S = 2\pi \lambda k_B T_s/\hbar$~\cite{ma2020atomistic}.

\begin{figure}[!ht]
\centering
\includegraphics[width=0.95\columnwidth]{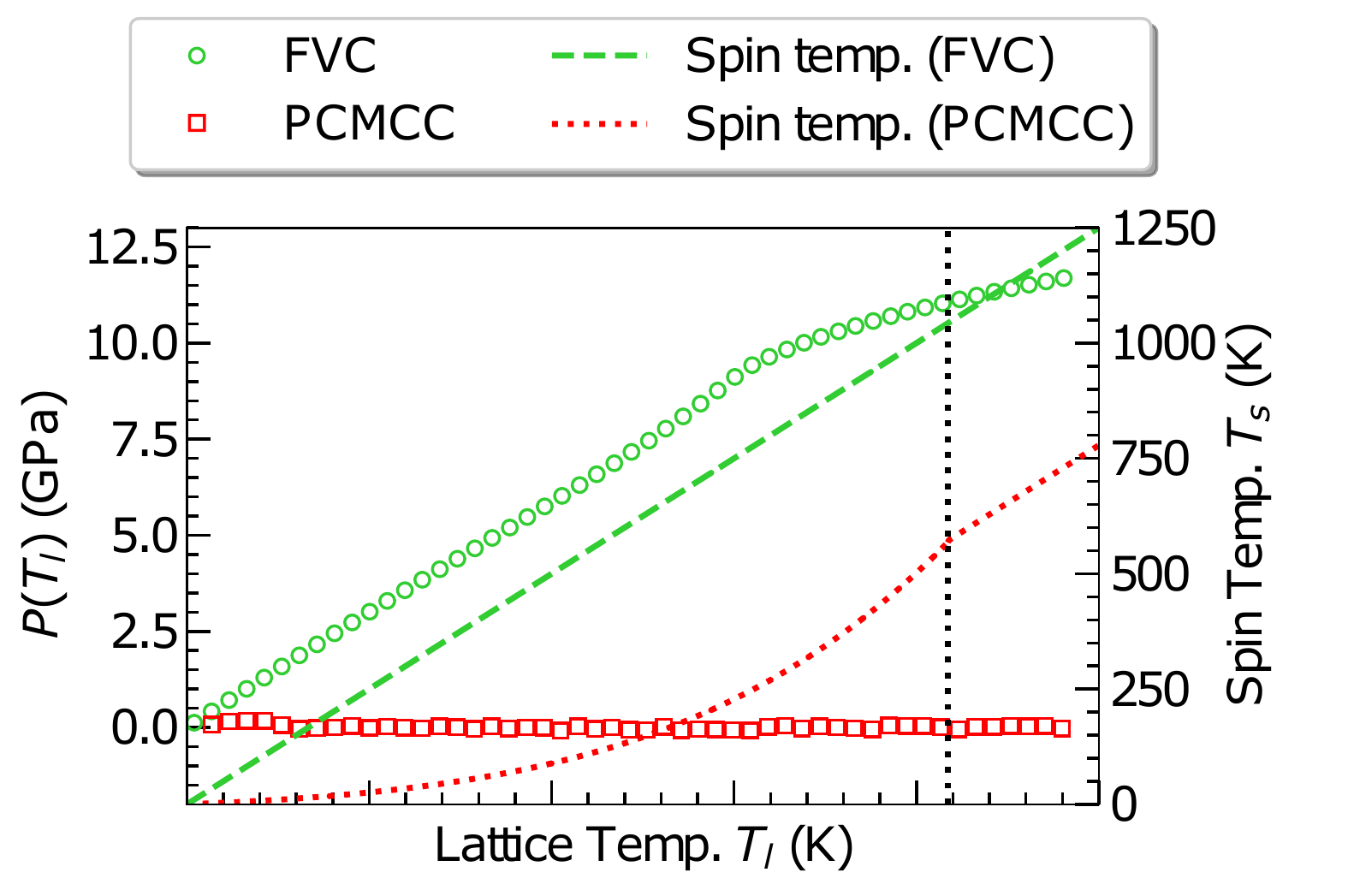}
\caption{
Symbols ((\markerone{}) and (\markerthree{})) associated to the left axis represent the pressure evolution in the FVC and PCMCC (similar to the PCC), as a function of lattice temperature. 
Doted lines associated to the right axis represent the spin thermostat temperature for the PCMCC and FVC (similar to PCC) as a function of the lattice temperature.
}
\label{fig:pressure}
\end{figure}

SD-MD calculations are carried out using a 20x20x20 BCC cell. The BCC lattice is oriented along each of the coordinate directions. The MD timestep in all cases is set to 0.1 femtoseconds. 
The damping constants are set to 0.1 (Gilbert damping, no units) for the spin thermostat, and to 0.1 picoseconds for the lattice thermostat.
Initially all spins start out aligned in the z-direction. 
To measure the magnetic properties for the canonical ensemble we initially thermalize the system under NVT dynamics at the target spin/lattice temperatures for 40 picoseconds and then sample the target properties for 10 picoseconds. 
For pressure-controlled simulations (see PCC and MCPCC in the "Results" section), after the initial 40 picosecond of temperature equilibration we freeze the spin configuration and run isobaric-isothermal NPT dynamics in order to allow the system to thermally expand (still accounting for the effect of the "magnetic" pressure, generated by the spin Hamiltonian). 
The pressure damping parameter is set to 10 picoseconds. The pressure equilibration run is terminated once the system pressure drops below 0.05 GPa. 
After this, the spin configuration is unfrozen and another equilibration run is carried out under NVT dynamics for 20 picoseconds. 
Unfreezing the spin configurations causes a small jump in the pressure, typically within the range of +/- 2 GPa. To reduce this pressure fluctuation, a series of uniform isotropic box deformations are performed under the NVE ensemble. During this procedure the box is deformed in 0.02\% increments every 2 picoseconds until the magnitude of the pressure is reduced to negligible values (< 10 MPa). 
Figure~\ref{fig:pressure} displays the pressure profiles obtained within the FVC and PCMCC (similar to the PCC).

For the magnetization-controlled conditions (PCMCC in the "Results" section), the spin temperature is adjusted to match the experimental magnetization values. Spin temperature adjustments are made based on the magnetization curve obtained in the pressure-controlled conditions (PCC in the "Results" section). The corresponding spin-lattice temperature relationship is shown in Eqs.~(\ref{spin-lattice_1}-\ref{spin-lattice_4}). Here the fitting coefficients are given as $a_{1} = 471.6$, $a_{2} = 6362$, $a_{3} = 2774$, $a_{4} = 1119$, $a_{5} = 13.6$, $a_{6} = 1043.3$, and $a_{7} = 0.1$, respectively. The functions $T_{s,pre}$ and $T_{s,post}$ prescribe how the spin temperature varies before and after the critical point. At the critical point we use a switching function $f_{sw}$ to smoothly transition from $T_{s,pre}$ to $T_{s,post}$:

\begin{eqnarray}
  T_{s,post}(T_{l}) &=& T_{l} - a_{1}
\label{spin-lattice_1}\\
  T_{s,pre}(T_{l}) &=& a_{2}\exp\left[-\left(\frac{T_{l} - a_{3}}{a_{4}} \right)^2\right] - a_{5}
\label{spin-lattice_2}\\
  f_{sw}(T_{l}) &=& \frac{1}{2} \left[ 1+ \tanh\left( \frac{T_{l} - a_{6}}{a_{7}}\right) \right]
\label{spin-lattice_3}\\
  T_{s}(T_{l}) &=& f_{sw}T_{s,post} + (1-f_{sw})T_{s,pre}
\label{spin-lattice_4}
\end{eqnarray}

Figure~\ref{fig:pressure} displays the spin temperature profiles for the FVC (and, similarly, the PCC), and the PCMCC.
After the magnetic measurements we compute elastic constants by performing both uniaxial and shear deformations along each of the coordinate directions and planes. The magnitude of these deformations in all cases is $2\%$ of the box length. Following each deformation the box is relaxed for 3 picoseconds. After this relaxation the stresses are sampled for 2 picoseconds.

\section*{Data Availability}
The data that support the findings of this study are available from the corresponding author upon reasonable request.

\section*{Code Availability}
The code which was used to train the SNAP potential is available from:
\url{https://github.com/FitSNAP/FitSNAP}.

\bibliography{main}

\section*{Acknowledgements}

All authors thank Mark Wilson for his detailed review and edits. 
Sandia National Laboratories is a multimission laboratory managed and operated by National Technology \& Engineering Solutions of Sandia, LLC, a wholly owned subsidiary of Honeywell International Inc., for the U.S. Department of Energy’s National Nuclear Security Administration under contract DE-NA0003525.
This paper describes objective technical results and analysis.
Any subjective views or opinions that might be expressed in the paper do not necessarily represent the views of the U.S. Department of Energy or the United States Government.
AC acknowledges funding from the Center for Advanced Systems Understanding (CASUS) which is financed by the German Federal Ministry of Education and Research (BMBF) and by the Saxon State Ministry for Science, Art, and Tourism (SMWK) with tax funds on the basis of the budget approved by the Saxon State Parliament.

\section*{Author contributions statement}

AC, MAW, MPD and JT performed the DFT calculations. 
JBM, MCM, JT and MAW generated the Database of configurations. 
JT implemented the extended spin Hamiltonian and the magnetic pressure computation in LAMMPS, and parametrized it on \emph{first-principles} calculations.
MAW and SN trained the SNAP potential. 
JT and SN performed the magneto-static calculations.
SN, APT, MAW and JT performed the magneto-dynamics calculations.
All authors participated in conceiving the research and writing the manuscript.

\section*{Competing interests}
The authors declare no competing interests.

\end{document}

%% file: table_db.tex
\begin{table*}[ht]
\begin{center}
{
\resizebox{\textwidth}{!}{
\begin{tabular}{lccccccc}
\hline\hline
 ~&$\#$ of Config. & $\#$ of Forces & Target Property & Energy Fit  Weight & Forces Fit Weight & Energy MAE (eV) & Forces MAE (eV$\cdot$\AA$^{-1}$) \\
\hline
  Eq. of State & 403 & 65286 & \makecell{Volumetric Deform} & $4.2\cdot10^{3}$& $2.0\cdot10^{5}$ & $1.6\cdot10^{-2}$ & $2.4\cdot10^{-1}$\\
  DFT-MD, 300K & 40 & 15360 & Bulk phonons & $2.9\cdot10^{5}$ & $1.1\cdot10^{5}$ & $5.2\cdot10^{-4}$ & $2.4\cdot10^{-1}$\\
  Liquid w/ Spins & 10 & 3000 & \makecell{Magnetic Disorder} & $5.5\cdot10^{1}$ & $1.9\cdot10^{4}$ & $2.0\cdot10^{-1}$ & $5.9\cdot10^{-1}$ \\
  Liquid w/o Spins & 52 & 15300 & \makecell{Structural Disorder} & $3.3\cdot10^{3}$ & $2.0\cdot10^{4}$ & $2.2\cdot10^{-1}$ & $8.0\cdot10^{-1}$\\
  Point Defects & 10 & 3096 & \makecell{Defect Energetics} & $1.4\cdot10^{2}$ & $3.5\cdot10^{4}$ & $2.8\cdot10^{-2}$ & $1.1\cdot10^{-1}$\\
  \makecell{Martensitic Transform} & 168 & 1008 & $\alpha \to \epsilon$ & $4.0\cdot10^{2}$ & $2.3\cdot10^{3}$ & $9.2\cdot10^{-2}$ & $2.3\cdot10^{-1}$\\
\hline\hline
\end{tabular}}
}
  \caption {Training set for linear SNAP model adapted from Ref. [\cite{goryaeva2019towards}] to include explicit spin degrees of freedom. Regression of SNAP coefficients takes into account both configuration energies and forces from DFT, optimization of group weights is applied to either term independently. Weighted linear regression is carried out via reported optimal fit weights, values have already been scaled by the number of training points each group contributes.
  The two last columns report the obtained mean-absolute errors (MAEs) in eV per atom.
  \label{tab:groups} }
\end{center}
\end{table*}

%% file: table_dakota.tex
\begin{table}[ht]
\begin{center}
{
\renewcommand{\arraystretch}{1.5}
\begin{tabular}{c c c c c}
\hline\hline
  ~ & SNAP & Exp/DFT & Units & Error $\%$  \\
\hline
  c$_{11}$ & 243.25 & 239.55 & GPa & 1.54$\%$ \\
  c$_{12}$ & 135.65 & 138.1 & GPa & 1.77$\%$ \\
  c$_{44}$ & 118.73 & 120.75 & GPa & 1.67$\%$ \\
  Bulk modulus & 171.52 & 169.55 & GPa & 1.16$\%$ \\
  $0.5$(c$_{11}$-c$_{12}$) & 53.8 & 51.9 & GPa & 3.66$\%$ \\
  Poisson ratio & 0.358 & 0.36 & - & 1.10$\%$ \\
  bcc energy & -8.25 & -8.26 & eV & 0.02$\%$ \\
  bcc lat. const. & 2.838 & 2.83 & \AA & 0.30$\%$ \\
\hline\hline
\end{tabular}
}
  \caption {Objective functions of the DAKOTA optimization with ground truth values taken from the present DFT calculations(at zero Kelvin) or experiments\cite{adams2006elastic}. Percent error is used as the objective function to avoid artificial importance scaling based on units of the target property.
  \label{tab:dakota} }
\end{center}
\end{table}

%% file: main.bbl
\begin{thebibliography}{10}
\urlstyle{rm}
\expandafter\ifx\csname url\endcsname\relax
  \def\url#1{\texttt{#1}}\fi
\expandafter\ifx\csname urlprefix\endcsname\relax\def\urlprefix{URL }\fi
\expandafter\ifx\csname doiprefix\endcsname\relax\def\doiprefix{DOI: }\fi
\providecommand{\bibinfo}[2]{#2}
\providecommand{\eprint}[2][]{\url{#2}}

\bibitem{tatsumoto1959temperature}
\bibinfo{author}{Tatsumoto, E.} \& \bibinfo{author}{Okamoto, T.}
\newblock \bibinfo{journal}{\bibinfo{title}{Temperature dependence of the
  magnetostriction constants in iron and silicon iron}}.
\newblock {\emph{\JournalTitle{Journal of the Physical Society of Japan}}}
  \textbf{\bibinfo{volume}{14}}, \bibinfo{pages}{1588--1594}
  (\bibinfo{year}{1959}).

\bibitem{bahl2009effect}
\bibinfo{author}{Bahl, C. R.~H.} \& \bibinfo{author}{Nielsen, K.~K.}
\newblock \bibinfo{journal}{\bibinfo{title}{The effect of demagnetization on
  the magnetocaloric properties of gadolinium}}.
\newblock {\emph{\JournalTitle{Journal of Applied Physics}}}
  \textbf{\bibinfo{volume}{105}}, \bibinfo{pages}{013916}
  (\bibinfo{year}{2009}).

\bibitem{tavares2000magnetic}
\bibinfo{author}{Tavares, S.}, \bibinfo{author}{Fruchart, D.},
  \bibinfo{author}{Miraglia, S.} \& \bibinfo{author}{Laborie, D.}
\newblock \bibinfo{journal}{\bibinfo{title}{Magnetic properties of an aisi 420
  martensitic stainless steel}}.
\newblock {\emph{\JournalTitle{Journal of alloys and compounds}}}
  \textbf{\bibinfo{volume}{312}}, \bibinfo{pages}{307--314}
  (\bibinfo{year}{2000}).

\bibitem{huang2018mapping}
\bibinfo{author}{Huang, S.}, \bibinfo{author}{Holmstr{\"o}m, E.},
  \bibinfo{author}{Eriksson, O.} \& \bibinfo{author}{Vitos, L.}
\newblock \bibinfo{journal}{\bibinfo{title}{Mapping the magnetic transition
  temperatures for medium-and high-entropy alloys}}.
\newblock {\emph{\JournalTitle{Intermetallics}}} \textbf{\bibinfo{volume}{95}},
  \bibinfo{pages}{80--84} (\bibinfo{year}{2018}).

\bibitem{rao2020unveiling}
\bibinfo{author}{Rao, Z.} \emph{et~al.}
\newblock \bibinfo{journal}{\bibinfo{title}{Unveiling the mechanism of abnormal
  magnetic behavior of fenicomncu high-entropy alloys through a joint
  experimental-theoretical study}}.
\newblock {\emph{\JournalTitle{Physical Review Materials}}}
  \textbf{\bibinfo{volume}{4}}, \bibinfo{pages}{014402} (\bibinfo{year}{2020}).

\bibitem{jaime2017piezomagnetism}
\bibinfo{author}{Jaime, M.} \emph{et~al.}
\newblock \bibinfo{journal}{\bibinfo{title}{Piezomagnetism and magnetoelastic
  memory in uranium dioxide}}.
\newblock {\emph{\JournalTitle{Nature communications}}}
  \textbf{\bibinfo{volume}{8}}, \bibinfo{pages}{1--7} (\bibinfo{year}{2017}).

\bibitem{nussle2019dynamic}
\bibinfo{author}{Nussle, T.}, \bibinfo{author}{Thibaudeau, P.} \&
  \bibinfo{author}{Nicolis, S.}
\newblock \bibinfo{journal}{\bibinfo{title}{Dynamic magnetostriction for
  antiferromagnets}}.
\newblock {\emph{\JournalTitle{Physical Review B}}}
  \textbf{\bibinfo{volume}{100}}, \bibinfo{pages}{214428}
  (\bibinfo{year}{2019}).

\bibitem{lejman2019magnetoelastic}
\bibinfo{author}{Lejman, M.} \emph{et~al.}
\newblock \bibinfo{journal}{\bibinfo{title}{Magnetoelastic and magnetoelectric
  couplings across the antiferromagnetic transition in multiferroic bifeo 3}}.
\newblock {\emph{\JournalTitle{Physical Review B}}}
  \textbf{\bibinfo{volume}{99}}, \bibinfo{pages}{104103}
  (\bibinfo{year}{2019}).

\bibitem{patrick2020spin}
\bibinfo{author}{Patrick, C.~E.}, \bibinfo{author}{Marchant, G.~A.} \&
  \bibinfo{author}{Staunton, J.~B.}
\newblock \bibinfo{journal}{\bibinfo{title}{Spin orientation and
  magnetostriction of tb 1- x dy x fe 2 from first principles}}.
\newblock {\emph{\JournalTitle{Physical Review Applied}}}
  \textbf{\bibinfo{volume}{14}}, \bibinfo{pages}{014091}
  (\bibinfo{year}{2020}).

\bibitem{GMVC1986:materials}
\bibinfo{author}{Graham, R.}, \bibinfo{author}{Morosin, B.},
  \bibinfo{author}{Venturini, E.} \& \bibinfo{author}{Carr, M.}
\newblock \bibinfo{journal}{\bibinfo{title}{Materials {{Modification}} and
  {{Synthesis Under High Pressure Shock Compression}}}}.
\newblock {\emph{\JournalTitle{Annual Review of Materials Science}}}
  \textbf{\bibinfo{volume}{16}}, \bibinfo{pages}{315--341},
  \doiprefix\url{10.1146/annurev.ms.16.080186.001531} (\bibinfo{year}{1986}).

\bibitem{surh2016magnetostructural}
\bibinfo{author}{Surh, M.~P.}, \bibinfo{author}{Benedict, L.~X.} \&
  \bibinfo{author}{Sadigh, B.}
\newblock \bibinfo{journal}{\bibinfo{title}{Magnetostructural transition
  kinetics in shocked iron}}.
\newblock {\emph{\JournalTitle{Physical review letters}}}
  \textbf{\bibinfo{volume}{117}}, \bibinfo{pages}{085701}
  (\bibinfo{year}{2016}).

\bibitem{Moses_NIF}
\bibinfo{author}{Moses, E.~I.}, \bibinfo{author}{Boyd, R.~N.},
  \bibinfo{author}{Remington, B.~A.}, \bibinfo{author}{Keane, C.~J.} \&
  \bibinfo{author}{Al-Ayat, R.}
\newblock \bibinfo{journal}{\bibinfo{title}{The national ignition facility:
  Ushering in a new age for high energy density science}}.
\newblock {\emph{\JournalTitle{Physics of Plasmas}}}
  \textbf{\bibinfo{volume}{16}}, \bibinfo{pages}{041006},
  \doiprefix\url{10.1063/1.3116505} (\bibinfo{year}{2009}).

\bibitem{Tschentscher_2017}
\bibinfo{author}{Tschentscher, T.} \emph{et~al.}
\newblock \bibinfo{journal}{\bibinfo{title}{Photon beam transport and
  scientific instruments at the european xfel}}.
\newblock {\emph{\JournalTitle{Applied Sciences}}}
  \textbf{\bibinfo{volume}{7}}, \doiprefix\url{10.3390/app7060592}
  (\bibinfo{year}{2017}).

\bibitem{tan2014combined}
\bibinfo{author}{Tan, X.}, \bibinfo{author}{Chan, S.}, \bibinfo{author}{Han,
  K.} \& \bibinfo{author}{Xu, H.}
\newblock \bibinfo{journal}{\bibinfo{title}{Combined effects of magnetic
  interaction and domain wall pinning on the coercivity in a bulk nd 60 fe 30
  al 10 ferromagnet}}.
\newblock {\emph{\JournalTitle{Scientific reports}}}
  \textbf{\bibinfo{volume}{4}}, \bibinfo{pages}{6805} (\bibinfo{year}{2014}).

\bibitem{gracia2020multicaloric}
\bibinfo{author}{Gr{\`a}cia-Condal, A.} \emph{et~al.}
\newblock \bibinfo{journal}{\bibinfo{title}{Multicaloric effects in
  metamagnetic heusler ni-mn-in under uniaxial stress and magnetic field}}.
\newblock {\emph{\JournalTitle{Applied Physics Reviews}}}
  \textbf{\bibinfo{volume}{7}}, \bibinfo{pages}{041406} (\bibinfo{year}{2020}).

\bibitem{wallace1960specific}
\bibinfo{author}{Wallace, D.~C.}, \bibinfo{author}{Sidles, P.} \&
  \bibinfo{author}{Danielson, G.}
\newblock \bibinfo{journal}{\bibinfo{title}{Specific heat of high purity iron
  by a pulse heating method}}.
\newblock {\emph{\JournalTitle{Journal of applied physics}}}
  \textbf{\bibinfo{volume}{31}}, \bibinfo{pages}{168--176}
  (\bibinfo{year}{1960}).

\bibitem{touloukian1970thermophysical}
\bibinfo{author}{Touloukian, Y.} \& \bibinfo{author}{Buyco, E.}
\newblock \bibinfo{journal}{\bibinfo{title}{Thermophysical properties of
  matter, vol. 4, specific heat}}.
\newblock {\emph{\JournalTitle{IFI/Plenum, New York}}}  (\bibinfo{year}{1970}).

\bibitem{chandler1987introduction}
\bibinfo{author}{Chandler, D.}
\newblock \emph{\bibinfo{title}{Introduction to modern statistical mechanics}}
  (\bibinfo{year}{1987}).

\bibitem{Hor2012:future}
\bibinfo{author}{Horstemeyer, M.~F.}
\newblock \bibinfo{title}{The near {{Future}}: {{ICME}} for the {{Creation}} of
  {{New Materials}} and {{Structures}}}.
\newblock In \emph{\bibinfo{booktitle}{Integrated {{Computational Materials
  Engineering}} ({{ICME}}) for {{Metals}}}}, chap.~\bibinfo{chapter}{10},
  \bibinfo{pages}{410--423}, \doiprefix\url{10.1002/9781118342664.ch10}
  (\bibinfo{publisher}{{John Wiley \& Sons, Ltd}}, \bibinfo{year}{2012}).

\bibitem{van2020roadmap}
\bibinfo{author}{van~der Giessen, E.} \emph{et~al.}
\newblock \bibinfo{journal}{\bibinfo{title}{Roadmap on multiscale materials
  modeling}}.
\newblock {\emph{\JournalTitle{Modelling and Simulation in Materials Science
  and Engineering}}} \textbf{\bibinfo{volume}{28}}, \bibinfo{pages}{043001}
  (\bibinfo{year}{2020}).

\bibitem{AW1959:studies}
\bibinfo{author}{Alder, B.~J.} \& \bibinfo{author}{Wainwright, T.~E.}
\newblock \bibinfo{journal}{\bibinfo{title}{Studies in {{Molecular Dynamics}}.
  {{I}}. {{General Method}}}}.
\newblock {\emph{\JournalTitle{The Journal of Chemical Physics}}}
  \textbf{\bibinfo{volume}{31}}, \bibinfo{pages}{459--466},
  \doiprefix\url{10.1063/1.1730376} (\bibinfo{year}{1959}).

\bibitem{rapaport2004art}
\bibinfo{author}{Rapaport, D.~C.}
\newblock \emph{\bibinfo{title}{The art of molecular dynamics simulation}}
  (\bibinfo{publisher}{Cambridge university press}, \bibinfo{year}{2004}).

\bibitem{plimpton1995fast}
\bibinfo{author}{Plimpton, S.}
\newblock \bibinfo{journal}{\bibinfo{title}{Fast parallel algorithms for
  short-range molecular dynamics}}.
\newblock {\emph{\JournalTitle{Journal of computational physics}}}
  \textbf{\bibinfo{volume}{117}}, \bibinfo{pages}{1--19}
  (\bibinfo{year}{1995}).

\bibitem{voter2002extending}
\bibinfo{author}{Voter, A.~F.}, \bibinfo{author}{Montalenti, F.} \&
  \bibinfo{author}{Germann, T.~C.}
\newblock \bibinfo{journal}{\bibinfo{title}{Extending the time scale in
  atomistic simulation of materials}}.
\newblock {\emph{\JournalTitle{Annual Review of Materials Research}}}
  \textbf{\bibinfo{volume}{32}}, \bibinfo{pages}{321--346}
  (\bibinfo{year}{2002}).

\bibitem{zepeda2017probing}
\bibinfo{author}{Zepeda-Ruiz, L.~A.}, \bibinfo{author}{Stukowski, A.},
  \bibinfo{author}{Oppelstrup, T.} \& \bibinfo{author}{Bulatov, V.~V.}
\newblock \bibinfo{journal}{\bibinfo{title}{Probing the limits of metal
  plasticity with molecular dynamics simulations}}.
\newblock {\emph{\JournalTitle{Nature}}} \textbf{\bibinfo{volume}{550}},
  \bibinfo{pages}{492--495} (\bibinfo{year}{2017}).

\bibitem{HBC+2017:universal}
\bibinfo{author}{Huan, T.~D.} \emph{et~al.}
\newblock \bibinfo{journal}{\bibinfo{title}{A universal strategy for the
  creation of machine learning-based atomistic force fields}}.
\newblock {\emph{\JournalTitle{npj Computational Materials}}}
  \textbf{\bibinfo{volume}{3}}, \doiprefix\url{10.1038/s41524-017-0042-y}
  (\bibinfo{year}{2017}).

\bibitem{Smith2017}
\bibinfo{author}{Smith, J.~S.}, \bibinfo{author}{Isayev, O.} \&
  \bibinfo{author}{Roitberg, A.~E.}
\newblock \bibinfo{journal}{\bibinfo{title}{Ani-1: an extensible neural network
  potential with dft accuracy at force field computational cost}}.
\newblock {\emph{\JournalTitle{Chem. Sci.}}} \textbf{\bibinfo{volume}{8}},
  \bibinfo{pages}{3192--3203}, \doiprefix\url{10.1039/C6SC05720A}
  (\bibinfo{year}{2017}).

\bibitem{ZHW+2018:deep}
\bibinfo{author}{Zhang, L.}, \bibinfo{author}{Han, J.}, \bibinfo{author}{Wang,
  H.}, \bibinfo{author}{Car, R.} \& \bibinfo{author}{E, W.}
\newblock \bibinfo{journal}{\bibinfo{title}{Deep {{Potential Molecular
  Dynamics}}: {{A Scalable Model}} with the {{Accuracy}} of {{Quantum
  Mechanics}}}}.
\newblock {\emph{\JournalTitle{Phys. Rev. Lett.}}}
  \textbf{\bibinfo{volume}{120}}, \bibinfo{pages}{143001},
  \doiprefix\url{10.1103/PhysRevLett.120.143001} (\bibinfo{year}{2018}).

\bibitem{BPKC2010:gaussian}
\bibinfo{author}{Bart{\'o}k, A.~P.}, \bibinfo{author}{Payne, M.~C.},
  \bibinfo{author}{Kondor, R.} \& \bibinfo{author}{Cs{\'a}nyi, G.}
\newblock \bibinfo{journal}{\bibinfo{title}{Gaussian {{Approximation
  Potentials}}: {{The Accuracy}} of {{Quantum Mechanics}}, without the
  {{Electrons}}}}.
\newblock {\emph{\JournalTitle{Phys. Rev. Lett.}}}
  \textbf{\bibinfo{volume}{104}}, \bibinfo{pages}{136403},
  \doiprefix\url{10.1103/PhysRevLett.104.136403} (\bibinfo{year}{2010}).

\bibitem{JNG2014:general}
\bibinfo{author}{{Jaramillo-Botero}, A.}, \bibinfo{author}{Naserifar, S.} \&
  \bibinfo{author}{Goddard, W.~A.}
\newblock \bibinfo{journal}{\bibinfo{title}{General {{Multiobjective Force
  Field Optimization Framework}}, with {{Application}} to {{Reactive Force
  Fields}} for {{Silicon Carbide}}}}.
\newblock {\emph{\JournalTitle{Journal of Chemical Theory and Computation}}}
  \textbf{\bibinfo{volume}{10}}, \bibinfo{pages}{1426--1439},
  \doiprefix\url{10.1021/ct5001044} (\bibinfo{year}{2014}).

\bibitem{LSB2018:hierarchical}
\bibinfo{author}{Lubbers, N.}, \bibinfo{author}{Smith, J.~S.} \&
  \bibinfo{author}{Barros, K.}
\newblock \bibinfo{journal}{\bibinfo{title}{Hierarchical modeling of molecular
  energies using a deep neural network}}.
\newblock {\emph{\JournalTitle{The Journal of Chemical Physics}}}
  \textbf{\bibinfo{volume}{148}}, \bibinfo{pages}{241715},
  \doiprefix\url{10.1063/1.5011181} (\bibinfo{year}{2018}).

\bibitem{TST+2015:spectral}
\bibinfo{author}{Thompson, A.~P.}, \bibinfo{author}{Swiler, L.~P.},
  \bibinfo{author}{Trott, C.~R.}, \bibinfo{author}{Foiles, S.~M.} \&
  \bibinfo{author}{Tucker, G.~J.}
\newblock \bibinfo{journal}{\bibinfo{title}{Spectral neighbor analysis method
  for automated generation of quantum-accurate interatomic potentials}}.
\newblock {\emph{\JournalTitle{Journal of Computational Physics}}}
  \textbf{\bibinfo{volume}{285}}, \bibinfo{pages}{316--330},
  \doiprefix\url{10.1016/j.jcp.2014.12.018} (\bibinfo{year}{2015}).

\bibitem{KS1965:selfconsistent}
\bibinfo{author}{Kohn, W.} \& \bibinfo{author}{Sham, L.~J.}
\newblock \bibinfo{journal}{\bibinfo{title}{Self-{{Consistent Equations
  Including Exchange}} and {{Correlation Effects}}}}.
\newblock {\emph{\JournalTitle{Phys. Rev.}}} \textbf{\bibinfo{volume}{140}},
  \bibinfo{pages}{A1133--A1138}, \doiprefix\url{10.1103/PhysRev.140.A1133}
  (\bibinfo{year}{1965}).

\bibitem{li2020complex}
\bibinfo{author}{Li, X.-G.}, \bibinfo{author}{Chen, C.},
  \bibinfo{author}{Zheng, H.}, \bibinfo{author}{Zuo, Y.} \&
  \bibinfo{author}{Ong, S.~P.}
\newblock \bibinfo{journal}{\bibinfo{title}{Complex strengthening mechanisms in
  the nbmotaw multi-principal element alloy}}.
\newblock {\emph{\JournalTitle{npj Computational Materials}}}
  \textbf{\bibinfo{volume}{6}}, \bibinfo{pages}{1--10} (\bibinfo{year}{2020}).

\bibitem{cusentino2020suppression}
\bibinfo{author}{Cusentino, M.}, \bibinfo{author}{Wood, M.} \&
  \bibinfo{author}{Thompson, A.}
\newblock \bibinfo{journal}{\bibinfo{title}{Suppression of helium bubble
  nucleation in beryllium exposed tungsten surfaces}}.
\newblock {\emph{\JournalTitle{Nuclear Fusion}}} \textbf{\bibinfo{volume}{60}},
  \bibinfo{pages}{126018} (\bibinfo{year}{2020}).

\bibitem{dragoni2018achieving}
\bibinfo{author}{Dragoni, D.}, \bibinfo{author}{Daff, T.~D.},
  \bibinfo{author}{Cs{\'a}nyi, G.} \& \bibinfo{author}{Marzari, N.}
\newblock \bibinfo{journal}{\bibinfo{title}{Achieving dft accuracy with a
  machine-learning interatomic potential: Thermomechanics and defects in bcc
  ferromagnetic iron}}.
\newblock {\emph{\JournalTitle{Physical Review Materials}}}
  \textbf{\bibinfo{volume}{2}}, \bibinfo{pages}{013808} (\bibinfo{year}{2018}).

\bibitem{ma2008large}
\bibinfo{author}{Ma, P.-W.}, \bibinfo{author}{Woo, C.} \&
  \bibinfo{author}{Dudarev, S.}
\newblock \bibinfo{journal}{\bibinfo{title}{Large-scale simulation of the
  spin-lattice dynamics in ferromagnetic iron}}.
\newblock {\emph{\JournalTitle{Physical review B}}}
  \textbf{\bibinfo{volume}{78}}, \bibinfo{pages}{024434}
  (\bibinfo{year}{2008}).

\bibitem{ma2016spilady}
\bibinfo{author}{Ma, P.-W.}, \bibinfo{author}{Dudarev, S.} \&
  \bibinfo{author}{Woo, C.}
\newblock \bibinfo{journal}{\bibinfo{title}{Spilady: A parallel cpu and gpu
  code for spin--lattice magnetic molecular dynamics simulations}}.
\newblock {\emph{\JournalTitle{Computer Physics Communications}}}
  \textbf{\bibinfo{volume}{207}}, \bibinfo{pages}{350--361}
  (\bibinfo{year}{2016}).

\bibitem{ma2020atomistic}
\bibinfo{author}{Ma, P.-W.} \& \bibinfo{author}{Dudarev, S.}
\newblock \bibinfo{journal}{\bibinfo{title}{Atomistic spin-lattice dynamics}}.
\newblock {\emph{\JournalTitle{Handbook of Materials Modeling: Methods: Theory
  and Modeling}}} \bibinfo{pages}{1017--1035} (\bibinfo{year}{2020}).

\bibitem{tranchida2018massively}
\bibinfo{author}{Tranchida, J.}, \bibinfo{author}{Plimpton, S.},
  \bibinfo{author}{Thibaudeau, P.} \& \bibinfo{author}{Thompson, A.~P.}
\newblock \bibinfo{journal}{\bibinfo{title}{Massively parallel symplectic
  algorithm for coupled magnetic spin dynamics and molecular dynamics}}.
\newblock {\emph{\JournalTitle{Journal of Computational Physics}}}
  \textbf{\bibinfo{volume}{372}}, \bibinfo{pages}{406--425}
  (\bibinfo{year}{2018}).

\bibitem{dos2020size}
\bibinfo{author}{Dos~Santos, G.} \emph{et~al.}
\newblock \bibinfo{journal}{\bibinfo{title}{Size-and temperature-dependent
  magnetization of iron nanoclusters}}.
\newblock {\emph{\JournalTitle{Physical Review B}}}
  \textbf{\bibinfo{volume}{102}}, \bibinfo{pages}{184426}
  (\bibinfo{year}{2020}).

\bibitem{zhou2020atomistic}
\bibinfo{author}{Zhou, Y.}, \bibinfo{author}{Tranchida, J.},
  \bibinfo{author}{Ge, Y.}, \bibinfo{author}{Murthy, J.} \&
  \bibinfo{author}{Fisher, T.~S.}
\newblock \bibinfo{journal}{\bibinfo{title}{Atomistic simulation of phonon and
  magnon thermal transport across the ferromagnetic-paramagnetic transition}}.
\newblock {\emph{\JournalTitle{Physical Review B}}}
  \textbf{\bibinfo{volume}{101}}, \bibinfo{pages}{224303}
  (\bibinfo{year}{2020}).

\bibitem{ma2017dynamic}
\bibinfo{author}{Ma, P.-W.}, \bibinfo{author}{Dudarev, S.} \&
  \bibinfo{author}{Wr{\'o}bel, J.~S.}
\newblock \bibinfo{journal}{\bibinfo{title}{Dynamic simulation of structural
  phase transitions in magnetic iron}}.
\newblock {\emph{\JournalTitle{Physical Review B}}}
  \textbf{\bibinfo{volume}{96}}, \bibinfo{pages}{094418}
  (\bibinfo{year}{2017}).

\bibitem{eldred2006dakota}
\bibinfo{author}{Eldred, M.~S.} \emph{et~al.}
\newblock \bibinfo{title}{Dakota, a multilevel parallel object-oriented
  framework for design optimization, parameter estimation, uncertainty
  quantification, and sensitivity analysis}.
\newblock \bibinfo{type}{Tech. Rep.}, \bibinfo{institution}{Citeseer}
  (\bibinfo{year}{2006}).

\bibitem{thompson2015spectral}
\bibinfo{author}{Thompson, A.~P.}, \bibinfo{author}{Swiler, L.~P.},
  \bibinfo{author}{Trott, C.~R.}, \bibinfo{author}{Foiles, S.~M.} \&
  \bibinfo{author}{Tucker, G.~J.}
\newblock \bibinfo{journal}{\bibinfo{title}{Spectral neighbor analysis method
  for automated generation of quantum-accurate interatomic potentials}}.
\newblock {\emph{\JournalTitle{Journal of Computational Physics}}}
  \textbf{\bibinfo{volume}{285}}, \bibinfo{pages}{316--330}
  (\bibinfo{year}{2015}).

\bibitem{crangle1971magnetization}
\bibinfo{author}{Crangle, J.} \& \bibinfo{author}{Goodman, G.}
\newblock \bibinfo{journal}{\bibinfo{title}{The magnetization of pure iron and
  nickel}}.
\newblock {\emph{\JournalTitle{Proceedings of the Royal Society of London. A.
  Mathematical and Physical Sciences}}} \textbf{\bibinfo{volume}{321}},
  \bibinfo{pages}{477--491} (\bibinfo{year}{1971}).

\bibitem{seki2005lattice}
\bibinfo{author}{Seki, I.} \& \bibinfo{author}{Nagata, K.}
\newblock \bibinfo{journal}{\bibinfo{title}{Lattice constant of iron and
  austenite including its supersaturation phase of carbon}}.
\newblock {\emph{\JournalTitle{ISIJ international}}}
  \textbf{\bibinfo{volume}{45}}, \bibinfo{pages}{1789--1794}
  (\bibinfo{year}{2005}).

\bibitem{basinski1955lattice}
\bibinfo{author}{Basinski, Z.~S.}, \bibinfo{author}{Hume-Rothery, W.} \&
  \bibinfo{author}{Sutton, A.}
\newblock \bibinfo{journal}{\bibinfo{title}{The lattice expansion of iron}}.
\newblock {\emph{\JournalTitle{Proceedings of the Royal Society of London.
  Series A. Mathematical and Physical Sciences}}}
  \textbf{\bibinfo{volume}{229}}, \bibinfo{pages}{459--467}
  (\bibinfo{year}{1955}).

\bibitem{ashcroft2016solid}
\bibinfo{author}{Ashcroft, N.~W.}, \bibinfo{author}{Mermin, N.~D.} \&
  \bibinfo{author}{Wei, D.}
\newblock \emph{\bibinfo{title}{Solid State Physics}}
  (\bibinfo{publisher}{Cengage Learning Asia Pte Limited},
  \bibinfo{year}{2016}).

\bibitem{dragoni2015thermoelastic}
\bibinfo{author}{Dragoni, D.}, \bibinfo{author}{Ceresoli, D.} \&
  \bibinfo{author}{Marzari, N.}
\newblock \bibinfo{journal}{\bibinfo{title}{Thermoelastic properties of
  $\alpha$-iron from first-principles}}.
\newblock {\emph{\JournalTitle{Physical Review B}}}
  \textbf{\bibinfo{volume}{91}}, \bibinfo{pages}{104105}
  (\bibinfo{year}{2015}).

\bibitem{dragoni2018vibrational}
\bibinfo{author}{Dragoni, D.}, \bibinfo{author}{Ceresoli, D.} \&
  \bibinfo{author}{Marzari, N.}
\newblock \bibinfo{journal}{\bibinfo{title}{Vibrational and thermoelastic
  properties of bcc iron from selected eam potentials}}.
\newblock {\emph{\JournalTitle{Computational Materials Science}}}
  \textbf{\bibinfo{volume}{152}}, \bibinfo{pages}{99--106}
  (\bibinfo{year}{2018}).

\bibitem{kormann2008free}
\bibinfo{author}{K{\"o}rmann, F.} \emph{et~al.}
\newblock \bibinfo{journal}{\bibinfo{title}{Free energy of bcc iron: Integrated
  ab initio derivation of vibrational, electronic, and magnetic
  contributions}}.
\newblock {\emph{\JournalTitle{Physical Review B}}}
  \textbf{\bibinfo{volume}{78}}, \bibinfo{pages}{033102}
  (\bibinfo{year}{2008}).

\bibitem{arakawa2020quantum}
\bibinfo{author}{Arakawa, K.} \emph{et~al.}
\newblock \bibinfo{journal}{\bibinfo{title}{Quantum de-trapping and transport
  of heavy defects in tungsten}}.
\newblock {\emph{\JournalTitle{Nature materials}}}
  \textbf{\bibinfo{volume}{19}}, \bibinfo{pages}{508--511}
  (\bibinfo{year}{2020}).

\bibitem{drautz2004spin}
\bibinfo{author}{Drautz, R.} \& \bibinfo{author}{F{\"a}hnle, M.}
\newblock \bibinfo{journal}{\bibinfo{title}{Spin-cluster expansion:
  Parametrization of the general adiabatic magnetic energy surface with ab
  initio accuracy}}.
\newblock {\emph{\JournalTitle{Physical Review B}}}
  \textbf{\bibinfo{volume}{69}}, \bibinfo{pages}{104404}
  (\bibinfo{year}{2004}).

\bibitem{drautz2020atomic}
\bibinfo{author}{Drautz, R.}
\newblock \bibinfo{journal}{\bibinfo{title}{Atomic cluster expansion of scalar,
  vectorial, and tensorial properties including magnetism and charge
  transfer}}.
\newblock {\emph{\JournalTitle{Physical Review B}}}
  \textbf{\bibinfo{volume}{102}}, \bibinfo{pages}{024104}
  (\bibinfo{year}{2020}).

\bibitem{marinica2012irradiation}
\bibinfo{author}{Marinica, M.-C.}, \bibinfo{author}{Willaime, F.} \&
  \bibinfo{author}{Crocombette, J.-P.}
\newblock \bibinfo{journal}{\bibinfo{title}{Irradiation-induced formation of
  nanocrystallites with c 15 laves phase structure in bcc iron}}.
\newblock {\emph{\JournalTitle{Physical review letters}}}
  \textbf{\bibinfo{volume}{108}}, \bibinfo{pages}{025501}
  (\bibinfo{year}{2012}).

\bibitem{chapman2020effect}
\bibinfo{author}{Chapman, J.~B.}, \bibinfo{author}{Ma, P.-W.} \&
  \bibinfo{author}{Dudarev, S.~L.}
\newblock \bibinfo{journal}{\bibinfo{title}{Effect of non-heisenberg magnetic
  interactions on defects in ferromagnetic iron}}.
\newblock {\emph{\JournalTitle{Physical Review B}}}
  \textbf{\bibinfo{volume}{102}}, \bibinfo{pages}{224106}
  (\bibinfo{year}{2020}).

\bibitem{chapman2019dynamics}
\bibinfo{author}{Chapman, J.~B.}, \bibinfo{author}{Ma, P.-W.} \&
  \bibinfo{author}{Dudarev, S.~L.}
\newblock \bibinfo{journal}{\bibinfo{title}{Dynamics of magnetism in fe-cr
  alloys with cr clustering}}.
\newblock {\emph{\JournalTitle{Physical Review B}}}
  \textbf{\bibinfo{volume}{99}}, \bibinfo{pages}{184413}
  (\bibinfo{year}{2019}).

\bibitem{klaver2006magnetism}
\bibinfo{author}{Klaver, T.}, \bibinfo{author}{Drautz, R.} \&
  \bibinfo{author}{Finnis, M.}
\newblock \bibinfo{journal}{\bibinfo{title}{Magnetism and thermodynamics of
  defect-free fe-cr alloys}}.
\newblock {\emph{\JournalTitle{Physical Review B}}}
  \textbf{\bibinfo{volume}{74}}, \bibinfo{pages}{094435}
  (\bibinfo{year}{2006}).

\bibitem{kalantar2005direct}
\bibinfo{author}{Kalantar, D.} \emph{et~al.}
\newblock \bibinfo{journal}{\bibinfo{title}{Direct observation of the $\alpha$-
  $\varepsilon$ transition in shock-compressed iron via nanosecond x-ray
  diffraction}}.
\newblock {\emph{\JournalTitle{Physical review letters}}}
  \textbf{\bibinfo{volume}{95}}, \bibinfo{pages}{075502}
  (\bibinfo{year}{2005}).

\bibitem{woo2015quantum}
\bibinfo{author}{Woo, C.}, \bibinfo{author}{Wen, H.}, \bibinfo{author}{Semenov,
  A.}, \bibinfo{author}{Dudarev, S.} \& \bibinfo{author}{Ma, P.-W.}
\newblock \bibinfo{journal}{\bibinfo{title}{Quantum heat bath for spin-lattice
  dynamics}}.
\newblock {\emph{\JournalTitle{Physical Review B}}}
  \textbf{\bibinfo{volume}{91}}, \bibinfo{pages}{104306}
  (\bibinfo{year}{2015}).

\bibitem{bergqvist2018realistic}
\bibinfo{author}{Bergqvist, L.} \& \bibinfo{author}{Bergman, A.}
\newblock \bibinfo{journal}{\bibinfo{title}{Realistic finite temperature
  simulations of magnetic systems using quantum statistics}}.
\newblock {\emph{\JournalTitle{Physical Review Materials}}}
  \textbf{\bibinfo{volume}{2}}, \bibinfo{pages}{013802} (\bibinfo{year}{2018}).

\bibitem{novikov2020machine}
\bibinfo{author}{Novikov, I.}, \bibinfo{author}{Grabowski, B.},
  \bibinfo{author}{Kormann, F.} \& \bibinfo{author}{Shapeev, A.}
\newblock \bibinfo{journal}{\bibinfo{title}{Machine-learning interatomic
  potentials reproduce vibrational and magnetic degrees of freedom}}.
\newblock {\emph{\JournalTitle{arXiv preprint arXiv:2012.12763}}}
  (\bibinfo{year}{2020}).

\bibitem{KRESSE199615}
\bibinfo{author}{Kresse, G.} \& \bibinfo{author}{Furthm{\"u}ller, J.}
\newblock \bibinfo{journal}{\bibinfo{title}{Efficiency of ab-initio total
  energy calculations for metals and semiconductors using a plane-wave basis
  set}}.
\newblock {\emph{\JournalTitle{Computational Materials Science}}}
  \textbf{\bibinfo{volume}{6}}, \bibinfo{pages}{15 -- 50},
  \doiprefix\url{https://doi.org/10.1016/0927-0256(96)00008-0}
  (\bibinfo{year}{1996}).

\bibitem{PhysRevB.59.1758}
\bibinfo{author}{Kresse, G.} \& \bibinfo{author}{Joubert, D.}
\newblock \bibinfo{journal}{\bibinfo{title}{From ultrasoft pseudopotentials to
  the projector augmented-wave method}}.
\newblock {\emph{\JournalTitle{Phys. Rev. B}}} \textbf{\bibinfo{volume}{59}},
  \bibinfo{pages}{1758--1775}, \doiprefix\url{10.1103/PhysRevB.59.1758}
  (\bibinfo{year}{1999}).

\bibitem{perdew1996generalized}
\bibinfo{author}{Perdew, J.~P.}, \bibinfo{author}{Burke, K.} \&
  \bibinfo{author}{Ernzerhof, M.}
\newblock \bibinfo{journal}{\bibinfo{title}{Generalized gradient approximation
  made simple}}.
\newblock {\emph{\JournalTitle{Physical review letters}}}
  \textbf{\bibinfo{volume}{77}}, \bibinfo{pages}{3865} (\bibinfo{year}{1996}).

\bibitem{blochl1994projector}
\bibinfo{author}{Bl{\"o}chl, P.~E.}
\newblock \bibinfo{journal}{\bibinfo{title}{Projector augmented-wave method}}.
\newblock {\emph{\JournalTitle{Physical review B}}}
  \textbf{\bibinfo{volume}{50}}, \bibinfo{pages}{17953} (\bibinfo{year}{1994}).

\bibitem{zimmermann2019comparison}
\bibinfo{author}{Zimmermann, B.} \emph{et~al.}
\newblock \bibinfo{journal}{\bibinfo{title}{Comparison of first-principles
  methods to extract magnetic parameters in ultrathin films: Co/pt (111)}}.
\newblock {\emph{\JournalTitle{Physical Review B}}}
  \textbf{\bibinfo{volume}{99}}, \bibinfo{pages}{214426}
  (\bibinfo{year}{2019}).

\bibitem{halilov1997magnon}
\bibinfo{author}{Halilov, S.}, \bibinfo{author}{Perlov, A.},
  \bibinfo{author}{Oppeneer, P.} \& \bibinfo{author}{Eschrig, H.}
\newblock \bibinfo{journal}{\bibinfo{title}{Magnon spectrum and related
  finite-temperature magnetic properties: A first-principle approach}}.
\newblock {\emph{\JournalTitle{EPL (Europhysics Letters)}}}
  \textbf{\bibinfo{volume}{39}}, \bibinfo{pages}{91} (\bibinfo{year}{1997}).

\bibitem{kurz2004ab}
\bibinfo{author}{Kurz, P.}, \bibinfo{author}{F{\"o}rster, F.},
  \bibinfo{author}{Nordstr{\"o}m, L.}, \bibinfo{author}{Bihlmayer, G.} \&
  \bibinfo{author}{Bl{\"u}gel, S.}
\newblock \bibinfo{journal}{\bibinfo{title}{Ab initio treatment of noncollinear
  magnets with the full-potential linearized augmented plane wave method}}.
\newblock {\emph{\JournalTitle{Physical Review B}}}
  \textbf{\bibinfo{volume}{69}}, \bibinfo{pages}{024415}
  (\bibinfo{year}{2004}).

\bibitem{sandratskii1998noncollinear}
\bibinfo{author}{Sandratskii, L.}
\newblock \bibinfo{journal}{\bibinfo{title}{Noncollinear magnetism in
  itinerant-electron systems: theory and applications}}.
\newblock {\emph{\JournalTitle{Advances in Physics}}}
  \textbf{\bibinfo{volume}{47}}, \bibinfo{pages}{91--160}
  (\bibinfo{year}{1998}).

\bibitem{marsman2002broken}
\bibinfo{author}{Marsman, M.} \& \bibinfo{author}{Hafner, J.}
\newblock \bibinfo{journal}{\bibinfo{title}{Broken symmetries in the
  crystalline and magnetic structures of $\gamma$-iron}}.
\newblock {\emph{\JournalTitle{Physical Review B}}}
  \textbf{\bibinfo{volume}{66}}, \bibinfo{pages}{224409}
  (\bibinfo{year}{2002}).

\bibitem{rosengaard1997finite}
\bibinfo{author}{Rosengaard, N.} \& \bibinfo{author}{Johansson, B.}
\newblock \bibinfo{journal}{\bibinfo{title}{Finite-temperature study of
  itinerant ferromagnetism in fe, co, and ni}}.
\newblock {\emph{\JournalTitle{Physical Review B}}}
  \textbf{\bibinfo{volume}{55}}, \bibinfo{pages}{14975} (\bibinfo{year}{1997}).

\bibitem{szilva2013interatomic}
\bibinfo{author}{Szilva, A.} \emph{et~al.}
\newblock \bibinfo{journal}{\bibinfo{title}{Interatomic exchange interactions
  for finite-temperature magnetism and nonequilibrium spin dynamics}}.
\newblock {\emph{\JournalTitle{Physical review letters}}}
  \textbf{\bibinfo{volume}{111}}, \bibinfo{pages}{127204}
  (\bibinfo{year}{2013}).

\bibitem{kaneyoshi1992introduction}
\bibinfo{author}{Kaneyoshi, T.}
\newblock \emph{\bibinfo{title}{Introduction to amorphous magnets}}
  (\bibinfo{publisher}{World Scientific Publishing Company},
  \bibinfo{year}{1992}).

\bibitem{yosida1996theory}
\bibinfo{author}{Yosida, K.}, \bibinfo{author}{Mattis, D.~C.} \&
  \bibinfo{author}{Yosida, K.}
\newblock \emph{\bibinfo{title}{THEORY OF MAGNETISM.: Edition en anglais}},
  vol. \bibinfo{volume}{122} (\bibinfo{publisher}{Springer Science \& Business
  Media}, \bibinfo{year}{1996}).

\bibitem{pajda2001ab}
\bibinfo{author}{Pajda, M.}, \bibinfo{author}{Kudrnovsk{\`y}, J.},
  \bibinfo{author}{Turek, I.}, \bibinfo{author}{Drchal, V.} \&
  \bibinfo{author}{Bruno, P.}
\newblock \bibinfo{journal}{\bibinfo{title}{Ab initio calculations of exchange
  interactions, spin-wave stiffness constants, and curie temperatures of fe,
  co, and ni}}.
\newblock {\emph{\JournalTitle{Physical Review B}}}
  \textbf{\bibinfo{volume}{64}}, \bibinfo{pages}{174402}
  (\bibinfo{year}{2001}).

\bibitem{yang1980generalizations}
\bibinfo{author}{Yang, K.-H.} \& \bibinfo{author}{Hirschfelder, J.~O.}
\newblock \bibinfo{journal}{\bibinfo{title}{Generalizations of classical
  poisson brackets to include spin}}.
\newblock {\emph{\JournalTitle{Physical Review A}}}
  \textbf{\bibinfo{volume}{22}}, \bibinfo{pages}{1814} (\bibinfo{year}{1980}).

\bibitem{loong1984neutron}
\bibinfo{author}{Loong, C.-K.}, \bibinfo{author}{Carpenter, J.},
  \bibinfo{author}{Lynn, J.}, \bibinfo{author}{Robinson, R.} \&
  \bibinfo{author}{Mook, H.}
\newblock \bibinfo{journal}{\bibinfo{title}{Neutron scattering study of the
  magnetic excitations in ferromagnetic iron at high energy transfers}}.
\newblock {\emph{\JournalTitle{Journal of applied physics}}}
  \textbf{\bibinfo{volume}{55}}, \bibinfo{pages}{1895--1897}
  (\bibinfo{year}{1984}).

\bibitem{lynn1975temperature}
\bibinfo{author}{Lynn, J.}
\newblock \bibinfo{journal}{\bibinfo{title}{Temperature dependence of the
  magnetic excitations in iron}}.
\newblock {\emph{\JournalTitle{Physical Review B}}}
  \textbf{\bibinfo{volume}{11}}, \bibinfo{pages}{2624} (\bibinfo{year}{1975}).

\bibitem{leger1972pressure}
\bibinfo{author}{Leger, J.}, \bibinfo{author}{Loriers-Susse, C.} \&
  \bibinfo{author}{Vodar, B.}
\newblock \bibinfo{journal}{\bibinfo{title}{Pressure effect on the curie
  temperatures of transition metals and alloys}}.
\newblock {\emph{\JournalTitle{Physical Review B}}}
  \textbf{\bibinfo{volume}{6}}, \bibinfo{pages}{4250} (\bibinfo{year}{1972}).

\bibitem{moran2003ab}
\bibinfo{author}{Mor{\'a}n, S.}, \bibinfo{author}{Ederer, C.} \&
  \bibinfo{author}{F{\"a}hnle, M.}
\newblock \bibinfo{journal}{\bibinfo{title}{Ab initio electron theory for
  magnetism in fe: Pressure dependence of spin-wave energies, exchange
  parameters, and curie temperature}}.
\newblock {\emph{\JournalTitle{Physical Review B}}}
  \textbf{\bibinfo{volume}{67}}, \bibinfo{pages}{012407}
  (\bibinfo{year}{2003}).

\bibitem{kormann2009pressure}
\bibinfo{author}{K{\"o}rmann, F.}, \bibinfo{author}{Dick, A.},
  \bibinfo{author}{Hickel, T.} \& \bibinfo{author}{Neugebauer, J.}
\newblock \bibinfo{journal}{\bibinfo{title}{Pressure dependence of the curie
  temperature in bcc iron studied by ab initio simulations}}.
\newblock {\emph{\JournalTitle{Physical Review B}}}
  \textbf{\bibinfo{volume}{79}}, \bibinfo{pages}{184406}
  (\bibinfo{year}{2009}).

\bibitem{skomski2008simple}
\bibinfo{author}{Skomski, R.} \emph{et~al.}
\newblock \emph{\bibinfo{title}{Simple models of magnetism}}
  (\bibinfo{publisher}{Oxford University Press on Demand},
  \bibinfo{year}{2008}).

\bibitem{wood2018extending}
\bibinfo{author}{Wood, M.~A.} \& \bibinfo{author}{Thompson, A.~P.}
\newblock \bibinfo{journal}{\bibinfo{title}{Extending the accuracy of the snap
  interatomic potential form}}.
\newblock {\emph{\JournalTitle{The Journal of Chemical Physics}}}
  \textbf{\bibinfo{volume}{148}}, \bibinfo{pages}{241721}
  (\bibinfo{year}{2018}).

\bibitem{zuo2020performance}
\bibinfo{author}{Zuo, Y.} \emph{et~al.}
\newblock \bibinfo{journal}{\bibinfo{title}{Performance and cost assessment of
  machine learning interatomic potentials}}.
\newblock {\emph{\JournalTitle{The Journal of Physical Chemistry A}}}
  \textbf{\bibinfo{volume}{124}}, \bibinfo{pages}{731--745}
  (\bibinfo{year}{2020}).

\bibitem{wood2019data}
\bibinfo{author}{Wood, M.~A.}, \bibinfo{author}{Cusentino, M.~A.},
  \bibinfo{author}{Wirth, B.~D.} \& \bibinfo{author}{Thompson, A.~P.}
\newblock \bibinfo{journal}{\bibinfo{title}{Data-driven material models for
  atomistic simulation}}.
\newblock {\emph{\JournalTitle{Physical Review B}}}
  \textbf{\bibinfo{volume}{99}}, \bibinfo{pages}{184305}
  (\bibinfo{year}{2019}).

\bibitem{deng2019electrostatic}
\bibinfo{author}{Deng, Z.}, \bibinfo{author}{Chen, C.}, \bibinfo{author}{Li,
  X.-G.} \& \bibinfo{author}{Ong, S.~P.}
\newblock \bibinfo{journal}{\bibinfo{title}{An electrostatic spectral neighbor
  analysis potential for lithium nitride}}.
\newblock {\emph{\JournalTitle{npj Computational Materials}}}
  \textbf{\bibinfo{volume}{5}}, \bibinfo{pages}{1--8} (\bibinfo{year}{2019}).

\bibitem{goryaeva2019towards}
\bibinfo{author}{Goryaeva, A.~M.}, \bibinfo{author}{Maillet, J.-B.} \&
  \bibinfo{author}{Marinica, M.-C.}
\newblock \bibinfo{journal}{\bibinfo{title}{Towards better efficiency of
  interatomic linear machine learning potentials}}.
\newblock {\emph{\JournalTitle{Computational Materials Science}}}
  \textbf{\bibinfo{volume}{166}}, \bibinfo{pages}{200--209}
  (\bibinfo{year}{2019}).

\bibitem{adams2006elastic}
\bibinfo{author}{Adams, J.~J.}, \bibinfo{author}{Agosta, D.},
  \bibinfo{author}{Leisure, R.} \& \bibinfo{author}{Ledbetter, H.}
\newblock \bibinfo{journal}{\bibinfo{title}{Elastic constants of monocrystal
  iron from 3 to 500 k}}.
\newblock {\emph{\JournalTitle{Journal of applied physics}}}
  \textbf{\bibinfo{volume}{100}}, \bibinfo{pages}{113530}
  (\bibinfo{year}{2006}).

\end{thebibliography}
